\documentclass[aps,prb,preprint,groupedaddress]{revtex4}

\usepackage{longtable}
\usepackage{graphicx}
\usepackage{subfigure}
\usepackage{epsfig}
\usepackage{color}

\begin{document}

\title{The energy landscape of silicon systems and its description by force fields,
tight binding schemes, density functional methods and Quantum Monte Carlo methods}


\author{S. Alireza Ghasemi$^1$, Maximilian Amsler$^1$, Richard G. Hennig$^2$,
Shantanu Roy$^1$, Stefan Goedecker$^{1,*}$, C. J. Umrigar$^3$, Luigi Genovese$^4$,
Thomas J. Lenosky$^5$, Tetsuya Morishita$^6$, Kengo Nishio$^6$}

\address{$^1$ Department of Physics, Universit\"{a}t Basel,
Klingelbergstr. 82, 4056 Basel, Switzerland}

\address{$^2$ Department of Materials Science and Engineering,
Cornell University, Ithaca, NY 14850, USA}

\affiliation{$^3$ Laboratory of Atomic and Solid State
Physics, Cornell University, Ithaca, New York 14853, USA}

\affiliation{$^4$ European Synchrotron Radiation Facility, 6 rue Horowitz, 38043 Grenoble (France) }

\affiliation{$^5$ C8 Medisensors, Los Gatos, CA 95032}

\affiliation{$^6$ Research Institute for Computational Sciences (RICS),
National Institute of Advanced Industrial Science and Technology (AIST),
1-1-1 Umezono, Tsukuba, Ibaraki 305-8568, Japan }

\email[]{Stefan.Goedecker@unibas.ch}
\homepage[]{http://pages.unibas.ch/comphys/comphys/}

\date{\today}

\begin{abstract}
The accuracy of the energy landscape of silicon systems 
obtained from various density functional methods, a tight 
binding scheme and force fields is studied. Quantum Monte 
Carlo results serve as quasi exact reference values.  In 
addition to the well known accuracy of DFT methods for 
geometric ground states and metastable configurations we 
find that DFT methods give a similar accuracy for transition 
states and thus a good overall description of the energy 
landscape. On the other hand, force fields give a very poor 
description of the landscape that are in most cases too 
rugged and contain many fake local minima and saddle points 
or ones that have the wrong height.
\end{abstract}


\maketitle 

\section{Introduction}

In spite of the great progress in density functional
methods for treating large systems, it is at present not possible
to treat systems with more than about 1000 atoms in complex
simulations where forces and energies have to be evaluated many
times. This is for instance necessary in molecular dynamics simulations
where one has to follow the evolution of the system over long time
intervals or in global optimization methods for finding the ground
state geometry. In these kinds of situations faster and more approximate
methods such as force fields or tight binding schemes are widely used.
Because of its technological importance several widely used force fields exist for silicon
and large scale simulations, which are not feasible with density functional methods, are
frequently performed using these more approximate methods.

These force fields are typically fitted
to a data set of ground state structures, usually containing crystalline
structures and sometimes also non-periodic structures. An accurate
description of some ground state geometries is however not sufficient to ensure accurate
dynamical simulations. Dynamical properties such as diffusion
coefficients are related to other properties of the energy landscape
such as barrier heights. The distribution of barrier heights and other
properties of the silicon potential energy surface (PES) have been studied using forcefields~\cite{Valiquette}.
In this paper we study overall properties of the
energy landscape which are relevant in many different contexts.
We look in particular at the accuracy of the
barrier heights in the various schemes used for large scale simulations
of silicon systems. Since it is known that the
barrier heights relevant to chemical reactions are not very well described with standard density
functionals such as local density approximation (LDA)\cite{ParrYang} or
Perdew-Burke-Ernzerhof (PBE),~\cite{PBE} we benchmark some of the energy
barriers using accurate quantum Monte Carlo methods.
There are two forms of QMC methods that are used for electronic
structure calculations, the simpler variational Monte Carlo (VMC) and
the more sophisticated diffusion Monte Carlo (DMC).  In VMC, quantum
mechanical expectation values are calculated using Monte Carlo
techniques to evaluate the many-dimensional integrals.  The accuracy of
the results depends crucially on the quality of the trial wave
function.  The DMC removes most of the error in the trial
wavefunction.  DMC is a stochastic projector method that projects out
the ground state from the trial wavefunction using an integral form of
the imaginary-time Schr\"odinger equation.  For Fermionic systems, the
antisymmetry constraint leads to the Fermion-sign problem that is cured by
fixing the nodes of the projected state to be those of an approximate trial wavefunction.
The resulting fixed-node error is the main uncontrolled error in DMC.
Currently, a systematic improvement of the wavefunction by
optimization of an increasing number of variational parameters is the
most practical approach for reducing the fixed-node error.~\cite{Umrigar,Toulouse07,Toulouse08}

Force fields and other approximate methods are sometimes applied to systems
that are very different from the systems that were in the fitting database.
The question is therefore how reliable are force field based structure
predictions of complex structures such as defects, interfaces or clusters.
In most studies of such systems only one force field was used but
in some exceptionally careful studies, such as in the study of dislocation kinks
in silicon~\cite{Kinks} and a comparative study of silicon empirical interatomic 
potentials~\cite{Balamane}, the results of several force fields were compared
and significant discrepancies in the results obtained from different force fields were indeed found.
To shed light on the accuracy of force fields we examine the configurational density of
states obtained by various approximate schemes for silicon.  Ideally there
would be a one-to-one mapping between stable structures (local minima)
obtained using approximate and accurate methods.
Consequently the density of configurations per energy interval would be identical for
approximate and accurate methods.

\section{Methods}

In our study we have included the most common force fields for silicon,
namely the Tersoff force field,~\cite{Tersoff} the Stillinger-Weber force
field,~\cite{StiWeb} the environmental-dependent interaction potential (EDIP)
force field~\cite{EDIP} and the Lenosky force
field.  The Tersoff force field, which has infinite range,
was smoothly extrapolated to zero by a third order
polynomial using cutoff radii that are large enough to ensure a smooth behavior of
the potential. The cutoff values were 2.7 and 3.3 \AA $\:$ where the smaller value denotes the
radius where the polynomial takes over and the larger value the radius
where the polynomial interaction drops to zero.
As a representative for tight binding schemes we use the Lenosky
tight (LTB) binding scheme.~\cite{Lenosky}
The QMC calculations are performed using the CHAMP code developed by
Umrigar, Filippi and Toulouse. The 1s, 2s, and 2p electrons of Si are eliminated
using a Hartree-Fock pseudopotential.~\cite{Alfe} The trial wave function
consists of a sum of Slater determinants of single-particle orbitals
multiplied by a Jastrow factor.  The orbitals of the Slater
determinant are taken from a DFT calculation using GAMESS~\cite{GAMESS1,GAMESS2}
with the B3LYP exchange-correlation functional.  The
excitations included in the sum of Slater determinants are those with the largest weights in a
configuration interaction with single and double excitation (CISD) calculation.
Configuration state functions (CSFs), i.e., linear combinations of determinants that have the correct spatial and spin symmetries, are
used to reduce the number of variational parameters.
The Jastrow correlation function describes electron-electron,
electron-nuclear and electron-electron nuclear correlations.  The
Jastrow parameters and the CSF coefficients are optimized in
variational Monte Carlo (VMC) using a recently developed energy
minimization method.~\cite{Umrigar,Toulouse07,Toulouse08}  Finally, diffusion Monte Carlo (DMC)
calculations using the optimized trial wavefunction and a time step of
0.01~Ha$^{-1}$ determine the energies of the structures.  Most of the calculations
employed a single determinant, but to estimate
the size of the fixed-node error we performed, for some structures, VMC and DMC calculations with
trial wavefunctions containing an increasing number of Slater determinants up to 150.
Apart from the preparation of trial wave function for QMC calculations and DFT barrier 
heights within B3LYP, all other DFT calculations are performed with the 
BigDFT~\cite{BigDFT} package, a pseudopotential based~\cite{GTH,HGH} 
DFT code with a wavelet basis set. 
Wavelets are a systematic basis set and the basis size was chosen 
sufficiently large that energies were converged to better than to 1 mHa.

\section{Transition States}

According to transition state theory, the height and the shape of
saddle points determine the dynamical behavior of a system.~\cite{Vineyard57}
Not much effort has been made to assess the effect of approximate energy
landscapes on the dynamics of silicon systems. 
In order to investigate the quality of the silicon potential energy landscape
within various schemes we performed simulations to find
saddle points of small Si$_8$ clusters using DFT within the
local density approximation. We then compared the barrier
heights for these LDA configurations with the heights obtained from other DFT schemes,
namely with a generalized gradient approximation functional (PBE), a hybrid
functional (B3LYP), and accurate quantum Monte Carlo (QMC) methods.
The 8 atom silicon cluster was chosen because for this size, QMC is
computationally not too expensive.
In agreement with previous work~\cite{Kanai} we found that the saddle point
geometries are nearly identical within different DFT functionals.
This justifies the use of LDA geometries in all energy calculations.
We also compared DFT saddle point results
with the Lenosky tight-binding (TB) method.
Quantum Monte Carlo has demonstrated its ability to provide
accurate reaction barrier heights.
Calculated values agreed with the experimental
values to within the statistical error bar of $0.07$~eV,~\cite{Grossman} for some organic molecules and
to within $0.005$~eV (see Refs.~\onlinecite{Ceperley,Diedrich})
for the well-known exchange reaction $H+H_2\longrightarrow H_2+H$.
We will therefore consider in the following our QMC results as accurate reference values.

To generate our saddle point configurations, we started with the putative global minimum of
the Si$_8$ cluster~\cite{Hellmann} and using the improved dimer method~\cite{Heyden}
we found six saddle points (SP1 to SP6) which connect the global minimum state
to other ones or to itself -- the latter corresponds to the exchange of
two silicon atoms (SP1 and SP4). Furthermore, two more saddle
points (SP7, SP8) are obtained starting from the first low-lying isomer.
One of them (SP7) corresponds to the exchange of two atoms.
Finding the saddle points and the adjacent minima was done within LDA using
the BigDFT~\cite{BigDFT} package.

\begin{table*}
\caption{Comparison of the barrier height (BH) energies of eight saddle point (SP) configurations relative to the two neighboring minima
calculated with various DFT schemes
and QMC.  The saddle point and minima configurations are from LDA. The abbreviation for the various DFT functionals
are defined in the text. VMC and DMC stand for variational and
diffusion QMC, MD-VMC and MD-DMC for multi-determinant VMC and DMC.
The HOMO-LUMO gaps for the minima (HLGM) and for the saddle
points (HLGS) are calculated within local density approximations using
the BigDFT code.  The statistical errors are given in parentheses.
All energies are in electron volts.\label{tab:barrierheights}}
\begin{ruledtabular}
\begin{tabular}{ccccccccccc}
system    & HLGM & HLGS &  LDA  &  PBE  & B3LYP &     VMC   &    DMC    & MD-VMC    & MD-DMC   &   TB   \\
\hline
SP1-BH1/2 & 1.45 & 0.55 & 0.359 & 0.367 & 0.405 & 0.434(23) & 0.406(13) & 0.329(9)  & 0.338(8) & -0.046 \\
SP2-BH1   & 1.45 & 0.41 & 0.735 & 0.729 & 0.770 & 0.809(20) & 0.804(18) &           &          & -0.286 \\
SP2-BH2   & 0.22 &      & 0.077 & 0.094 & 0.058 & 0.103(22) & 0.069(18) &           &          &  0.166 \\
SP3-BH1   & 0.96 & 0.64 & 0.043 & 0.050 & 0.056 & 0.046(18) & 0.028(15) & 0.042(9)  & 0.028(8) &  0.201 \\
SP3-BH2   & 1.45 &      & 2.900 & 2.689 & 2.328 & 2.927(17) & 2.845(15) & 2.969(8)  & 2.826(8) &  1.934 \\
SP4-BH1/2 & 1.45 & 0.13 & 1.065 & 1.053 & 1.075 & 1.233(18) & 1.181(15) & 1.131(10) & 1.134(7) &  0.496 \\
SP5-BH1   & 1.21 & 0.42 & 0.338 & 0.346 & 0.324 & 0.410(18) & 0.378(14) &           &          &  0.398 \\
SP5-BH2   & 1.45 &      & 0.766 & 0.761 & 0.800 & 0.883(17) & 0.862(16) &           &          &  0.019 \\
SP6-BH1   & 1.49 & 0.69 & 0.581 & 0.573 & 0.514 & 0.503(18) & 0.536(16) &           &          & -0.082 \\
SP6-BH2   & 1.45 &      & 1.198 & 1.158 & 1.015 & 1.283(18) & 1.194(16) &           &          &  0.921 \\
SP7-BH1/2 & 1.12 & 0.58 & 0.289 & 0.272 & 0.192 & 0.228(18) & 0.224(16) &           &          &  0.157 \\
SP8-BH1   & 0.22 & 0.82 & 0.212 & 0.198 & 0.041 & 0.143(17) & 0.120(16) &           &          &  0.144 \\
SP8-BH2   & 1.12 &      & 0.445 & 0.420 & 0.406 & 0.491(18) & 0.445(16) &           &          & -0.264
\end{tabular}
\end{ruledtabular}
\end{table*}

\begin{figure*}            
\setlength{\unitlength}{1cm}
\subfigure[~minimum1]{\includegraphics[width=0.32\columnwidth,angle=0]{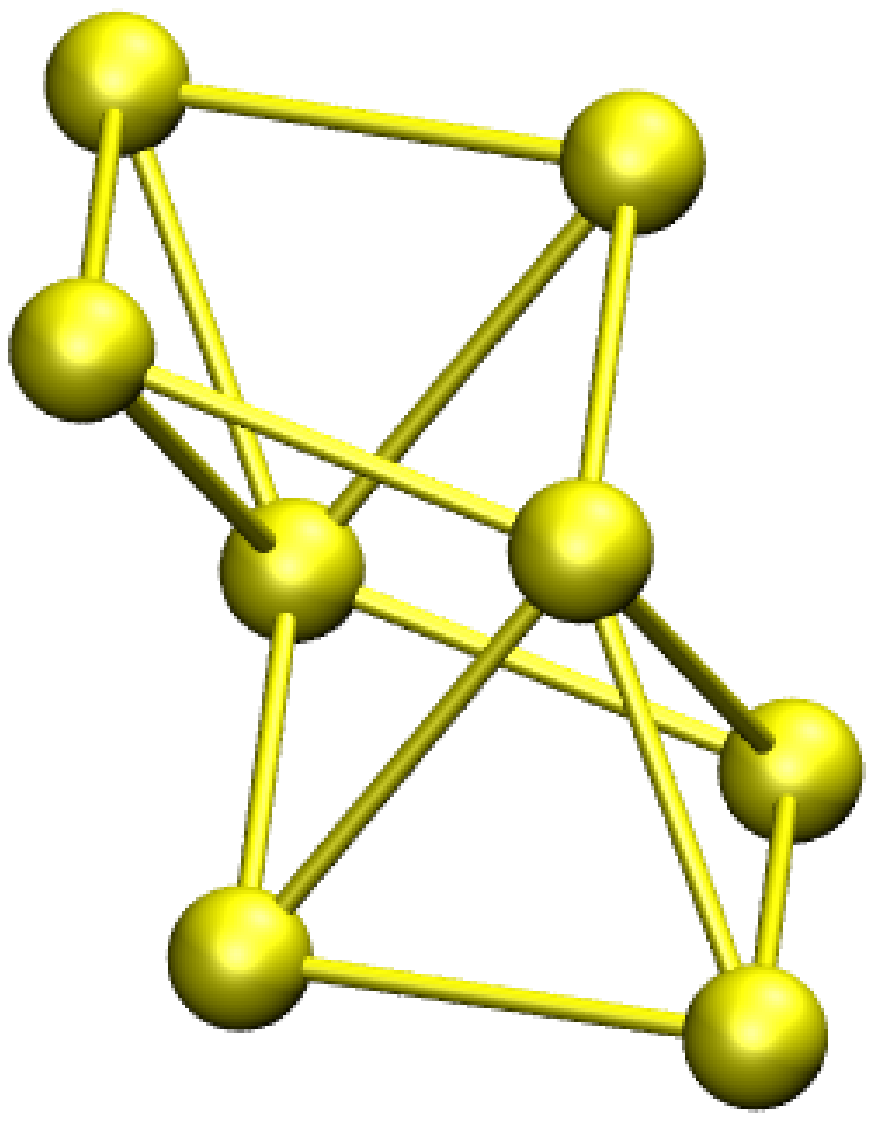}} 
\subfigure[~saddle  ]{\includegraphics[width=0.32\columnwidth,angle=0]{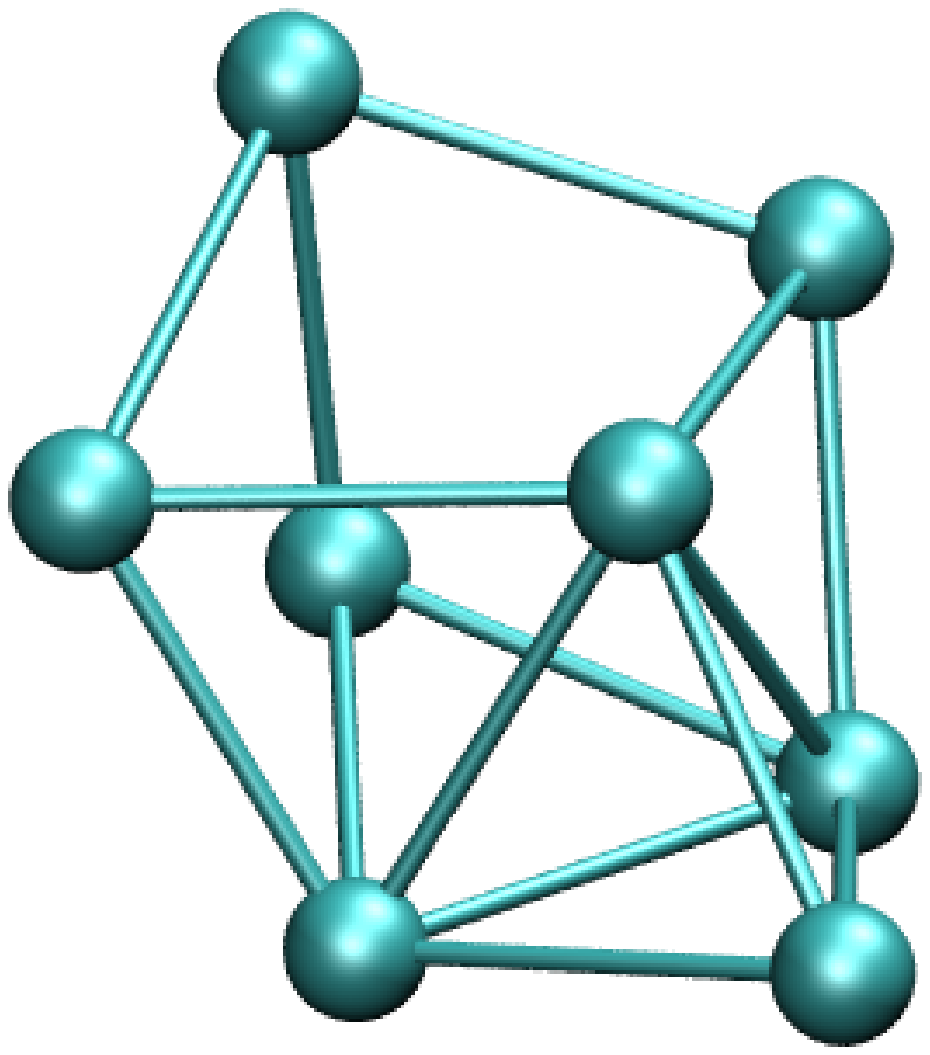}} 
\subfigure[~minimum2]{\includegraphics[width=0.32\columnwidth,angle=0]{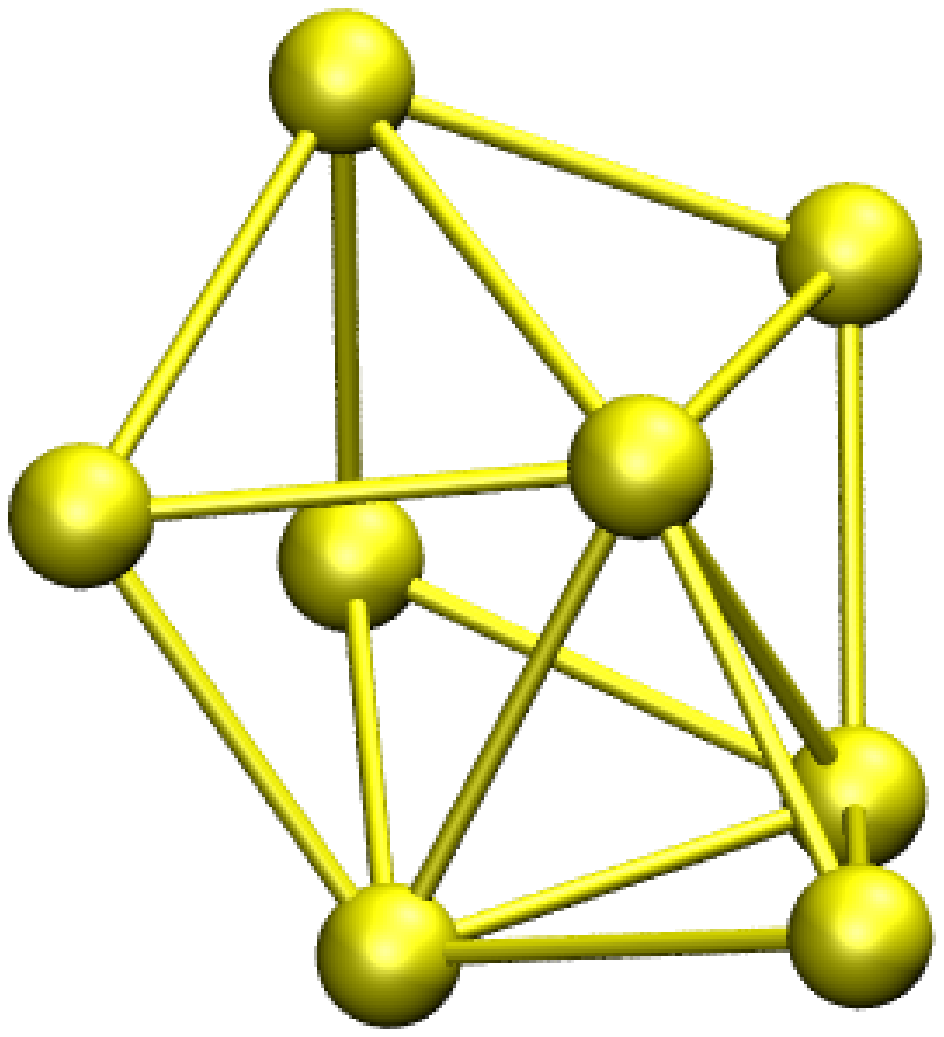}} 
\caption{\label{fig:SP2set} (color online) The saddle point SP2 (b) and the two neighboring
minima (a), (c). It is obvious that the silicon atoms in the saddle point and the minima
configurations are all in a similar environment, in particular the saddle point
is very similar to the structure (c).}
\end{figure*}

\begin{figure}            
\setlength{\unitlength}{1cm}
\subfigure[~SP1]{\includegraphics[width=0.35\columnwidth,angle=0]{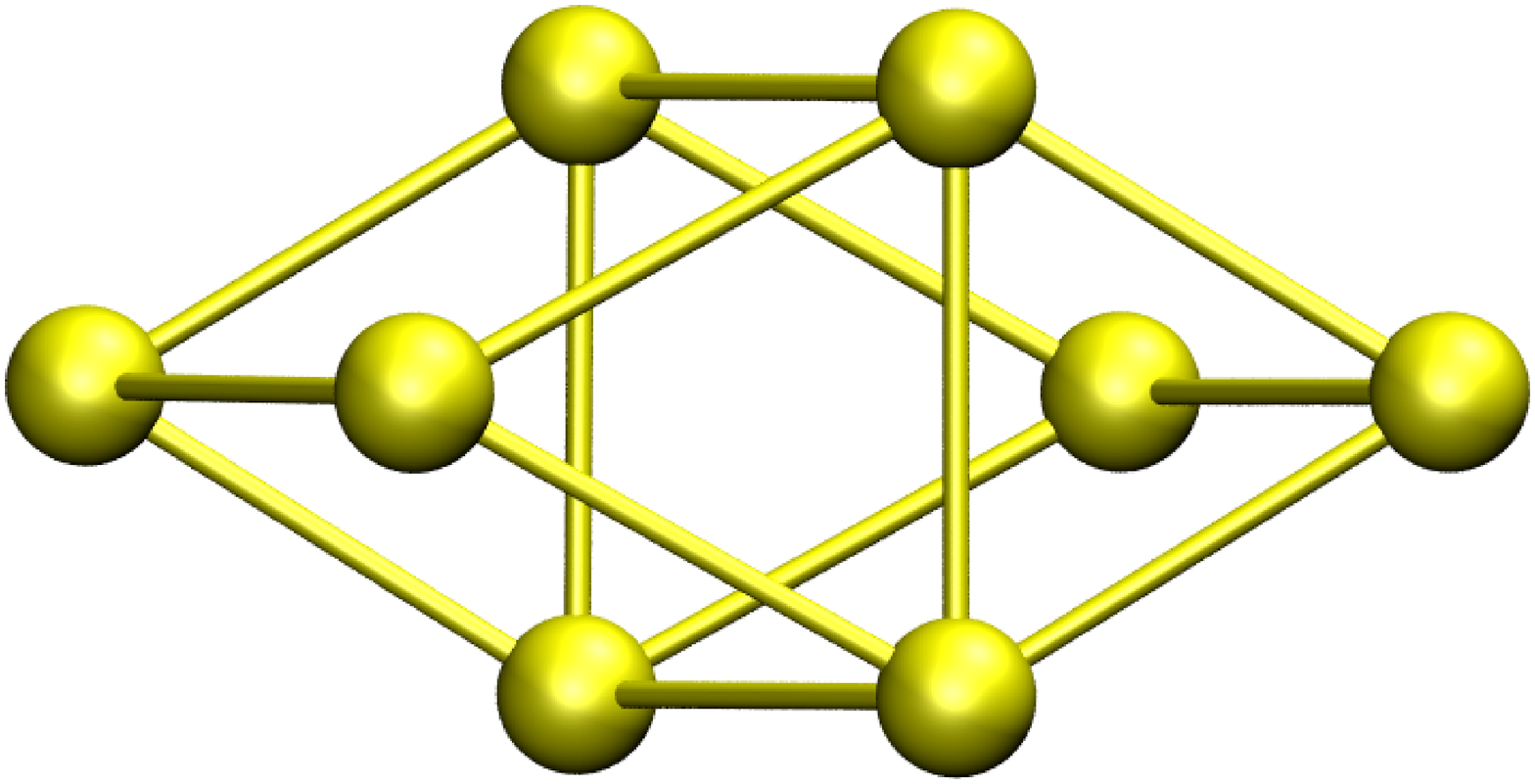}} 
\subfigure[~SP2]{\includegraphics[width=0.35\columnwidth,angle=0]{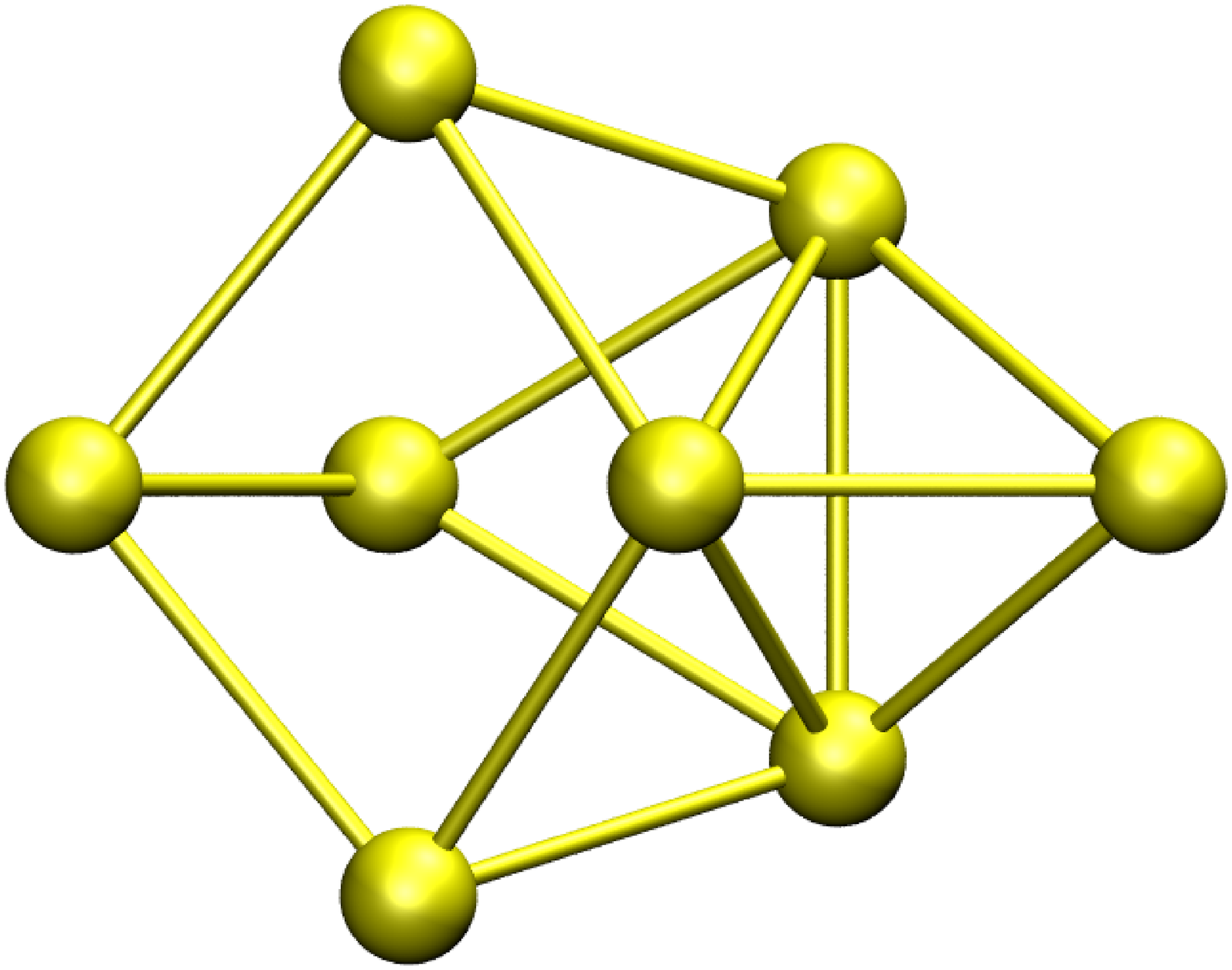}} 
\subfigure[~SP3]{\includegraphics[width=0.35\columnwidth,angle=0]{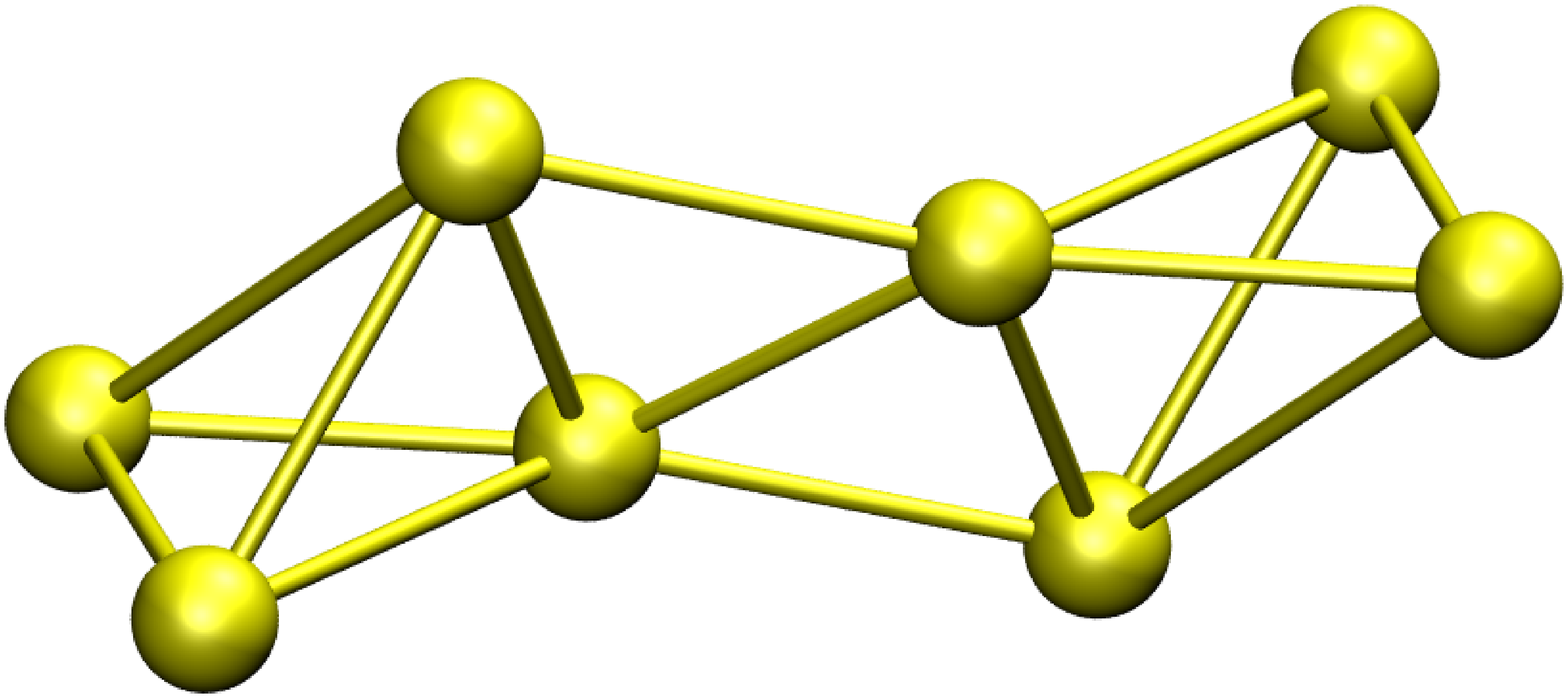}} 
\subfigure[~SP4]{\includegraphics[width=0.35\columnwidth,angle=0]{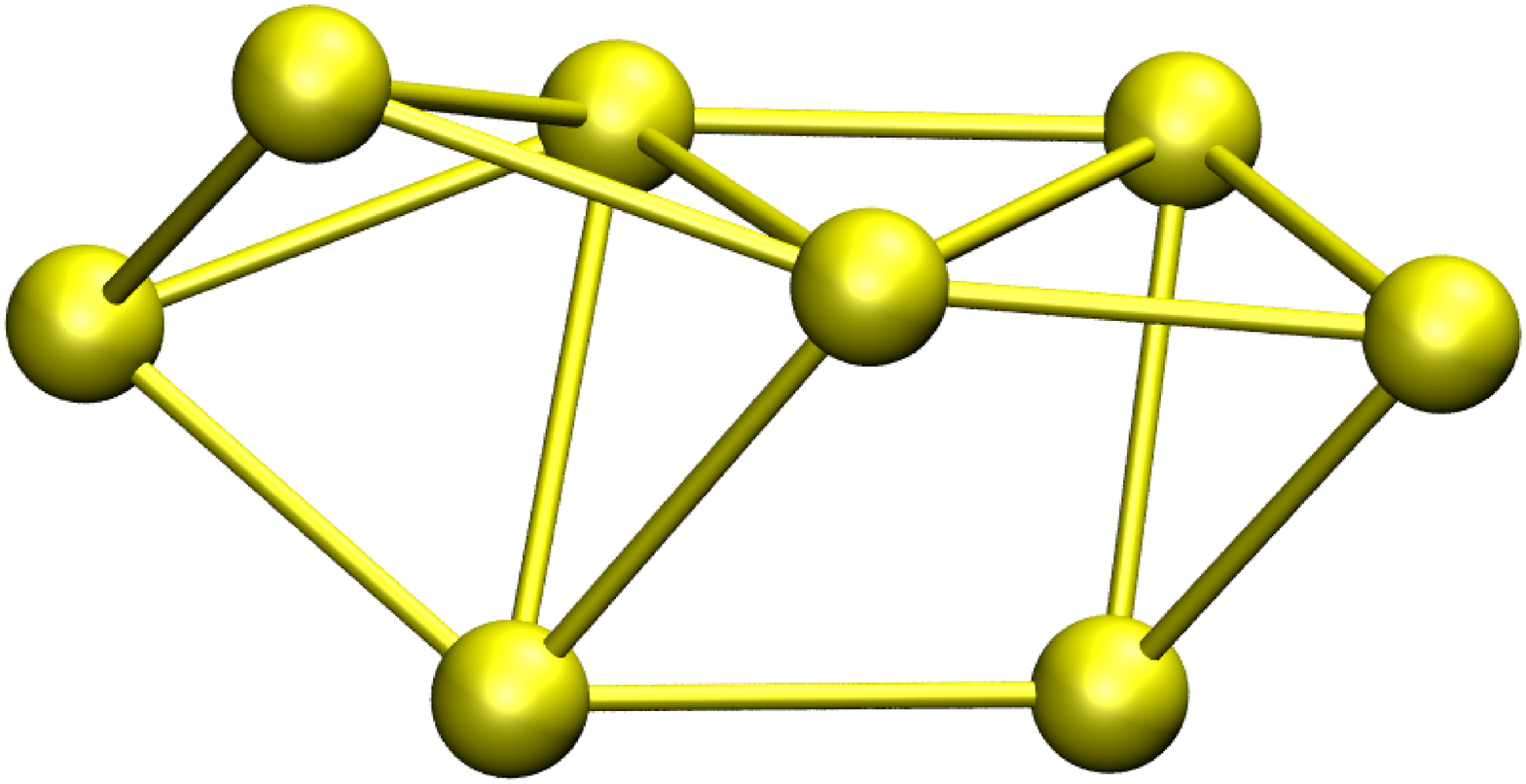}} 
\subfigure[~SP5]{\includegraphics[width=0.35\columnwidth,angle=0]{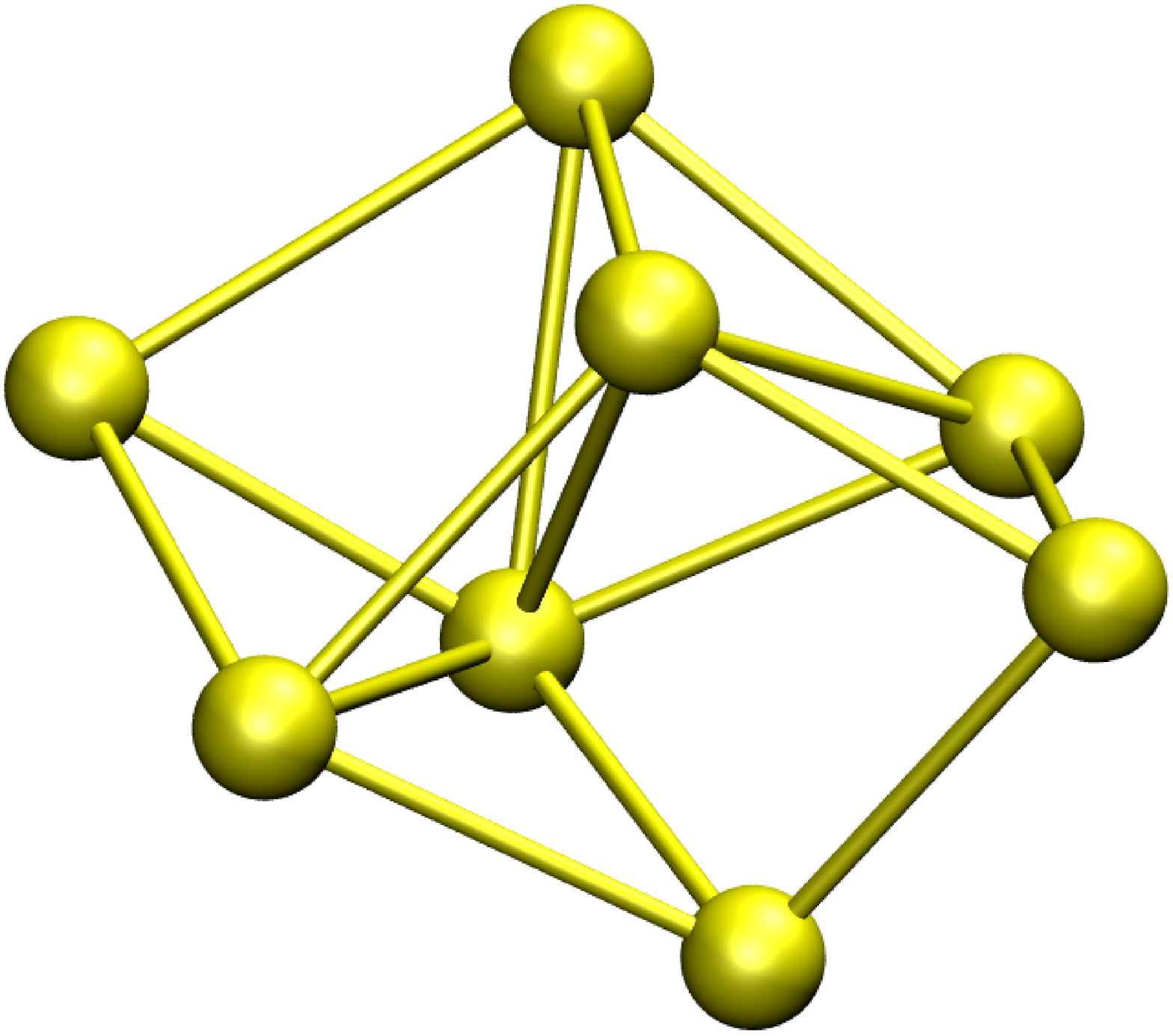}} 
\subfigure[~SP6]{\includegraphics[width=0.35\columnwidth,angle=0]{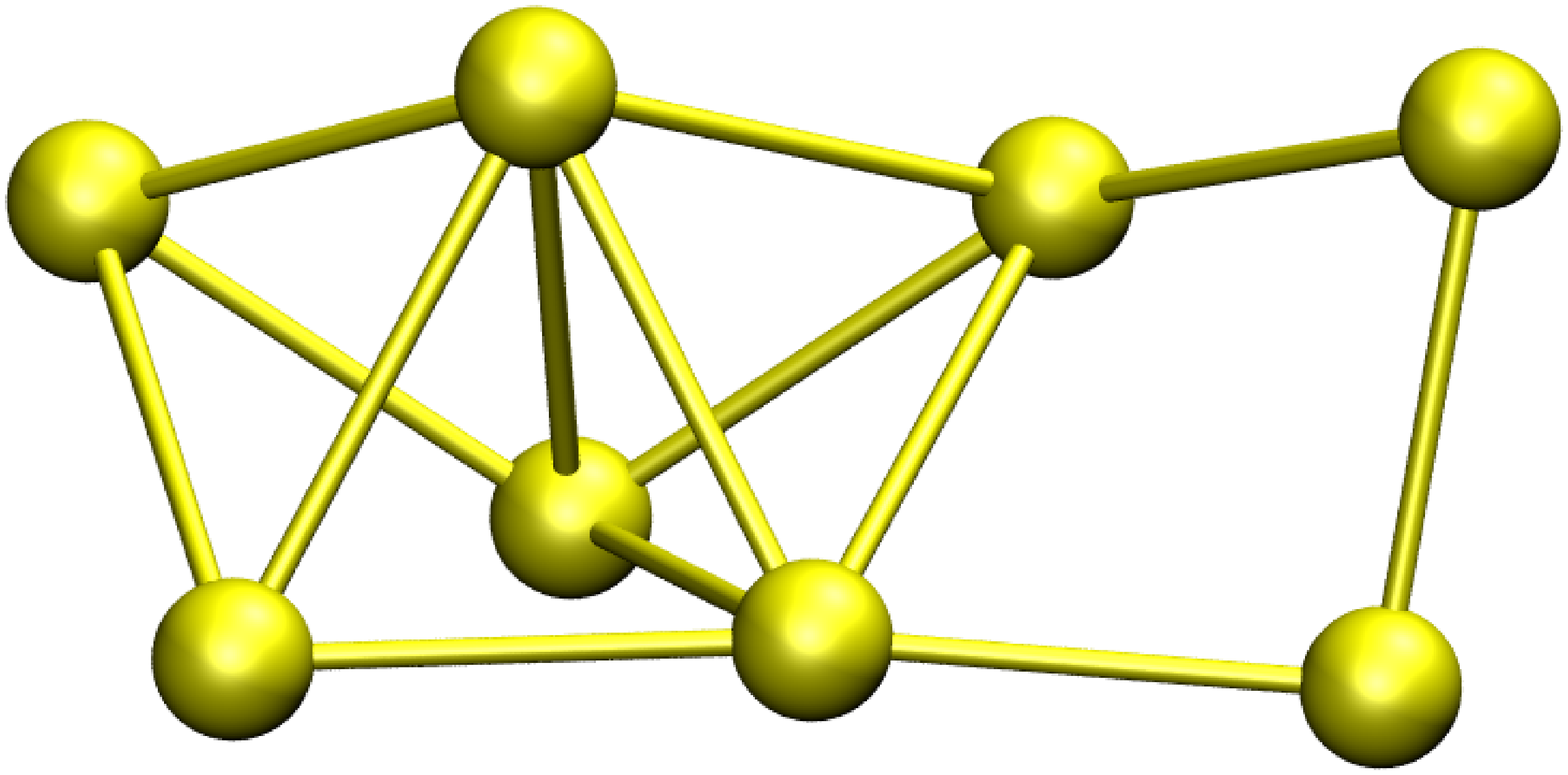}} 
\subfigure[~SP7]{\includegraphics[width=0.35\columnwidth,angle=0]{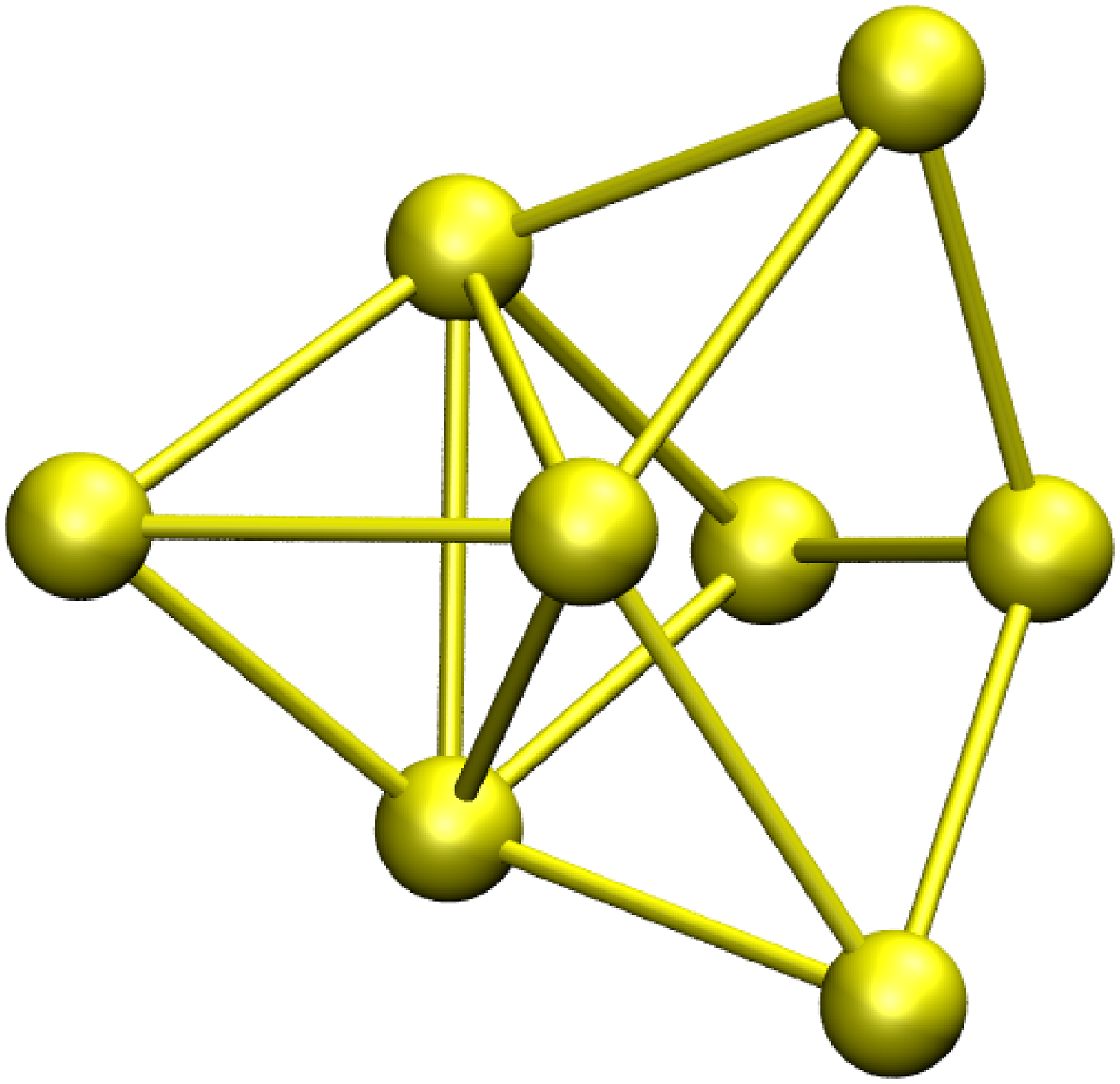}} 
\subfigure[~SP8]{\includegraphics[width=0.35\columnwidth,angle=0]{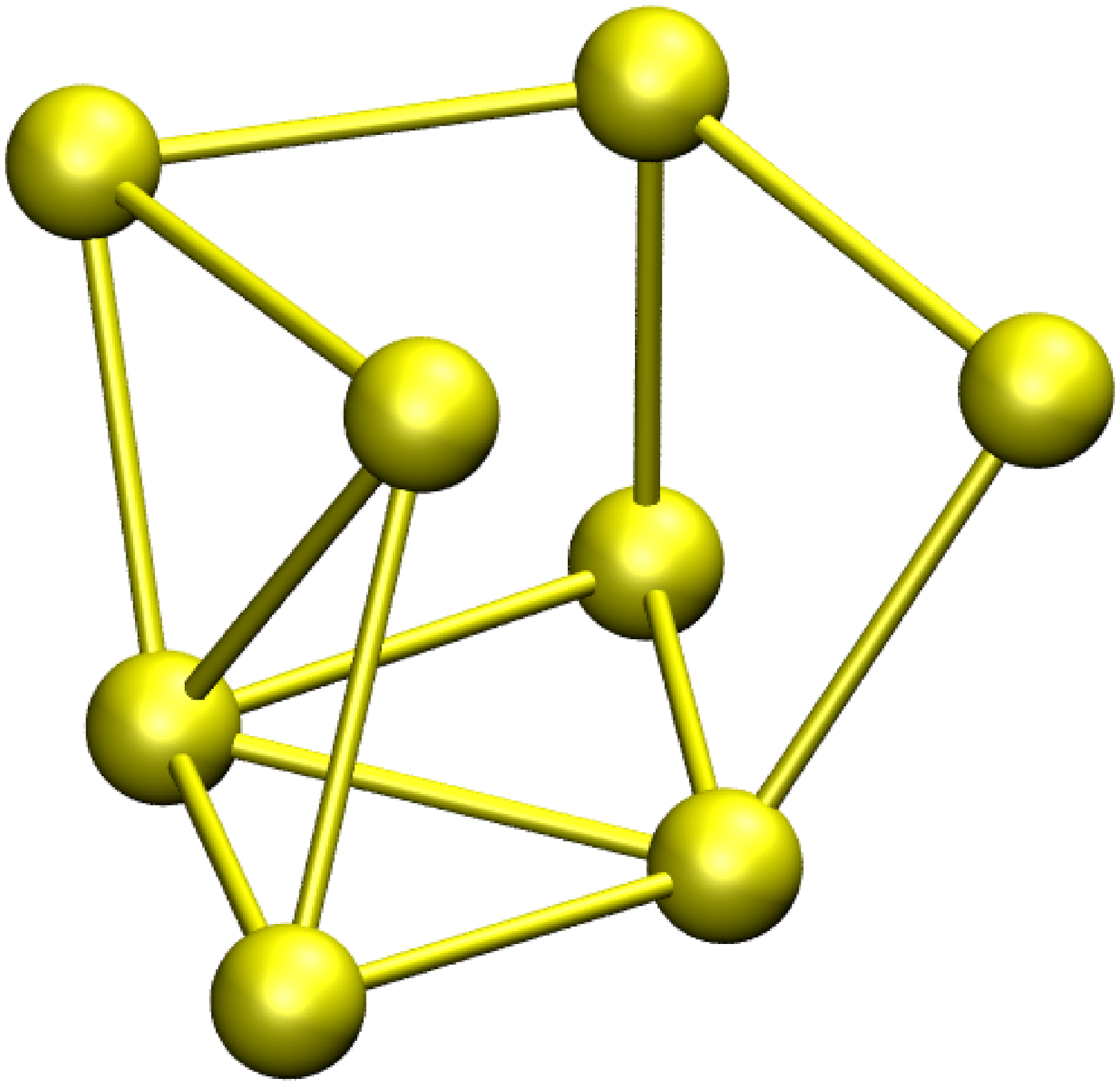}} 
\caption{\label{fig:saddlepoints} (color online) The eight Si$_8$ saddle points obtained by LDA calculations.}
\end{figure}

There are numerous publications (for a survey see Ref.~\onlinecite{Zhao}) in which it is shown that
DFT schemes give poor transition state barrier heights for
chemical reactions.~\cite{Andersson,Gruning,Patchkovskii} Within LDA and GGA's the results are most unsatisfactory for hydrogen
transfer reactions where covalent hydrogen bonds are broken and formed.
The most simple and prominent example is the exchange reaction
$H+H_2\longrightarrow H_2+H$
where the DFT schemes do not predict a barrier at all.
The poor performance seems to be due to poor cancellation of the electrostatic self-interaction errors
in the DFT schemes.~\cite{Perdew} In the literature this problem
is known as ``self-interaction error''
which is related to the delocalization error.~\cite{Cohen}
Hybrid functionals, which give a better error cancellation, give
improved barrier heights in this case.~\cite{Zhao}
Nevertheless researchers usually resort to wavefunction methods if highly accurate barrier heights
are needed for chemical reactions.

Table~\ref{tab:barrierheights} shows that in our case the situation is entirely different.
Already at the LDA level the barrier heights are reasonably accurate and actually slightly better than the B3LYP and PBE barriers.
How can this surprising accuracy be explained? In contrast to chemical reactions our clusters are never
torn apart into fragments when they move along the minimum energy pathway from one local
minimum over a saddle point into another local minimum.
Even at the transition state (see Figs.~\ref{fig:SP2set} and \ref{fig:saddlepoints})
the silicon atoms are all in an environment that is similar to the environment at a local minimum
and one cannot distinguish a saddle point configuration from a local minimum energy configuration
by inspection.
DFT self-interaction errors~\cite{Cohen} are therefore expected to cancel to a large degree.

\begin{figure}
\centering
\includegraphics[width=1.0\columnwidth]{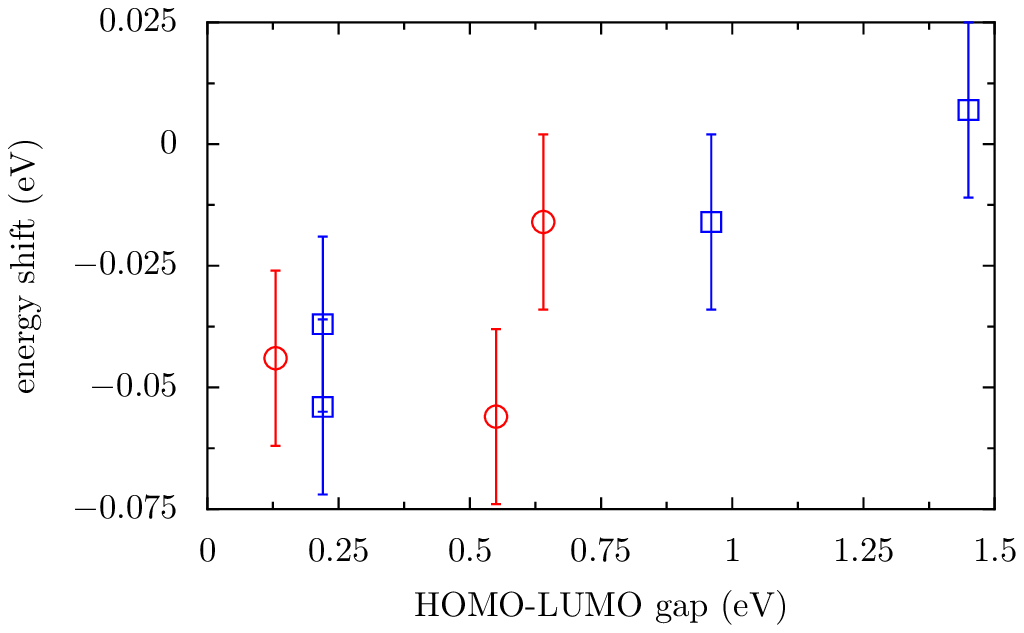}
\caption{(color online) The correlation between the energy change from single- to multi-determinant
diffusion Monte Carlo calculations with the HOMO-LUMO gap.
The (red) circles are for saddle point structures
and the (blue) squares are for minimum point structures.\label{fig:multideterminant}}
\end{figure}

Since DFT is essentially a one determinant method one would expect that DFT results are
particularly poor when transition states have multi-determinant character. This is frequently
the case in chemical reactions and under such circumstances multi-reference wavefunction methods have
to be employed if accurate barrier heights are needed.  Small HOMO-LUMO gaps are
an indication of the importance of multi-reference configurations.
The HOMO-LUMO gaps of all saddle points and the adjacent minima
are presented in Table~\ref{tab:barrierheights}.  The average
HOMO-LUMO gap of saddle points is less than that of the minima by $\approx 0.65$~eV.
As usual the LDA and GGA gaps are much smaller than the B3LYP gaps
which are certainly more accurate. For the configurations with small HOMO-LUMO
gaps we also did multi-determinant QMC calculations with as many as 150 determinants.
The DMC energies went down by no more than 2 mHa.
In Fig.~\ref{fig:multideterminant},
the correlation between the HOMO-LUMO gap and the energy change from single- to multi-determinant
calculations is illustrated for seven configurations, in particular for the three
configurations that have a HOMO-LUMO gap of less than $0.25$ eV.
The results show that the influence of a multi determinant wavefunction on the barrier height is very small.
Another indication that the multi-determinant character of the saddle point configurations can be neglected is
the fact the natural occupation numbers drop rapidly from one to zero. The occupation numbers obtained from
CISD calculations using GAMESS
with about 70 determinants are shown in Table~\ref{tab:occupationnumber}.

\begin{table*}
\caption{The occupation number of the four highest occupied orbitals and the
four lowest unoccupied orbitals of Si$_8$ clusters obtained from CISD 
calculations.\label{tab:occupationnumber}}
\begin{ruledtabular}
\begin{tabular}{ccccccccc}
 system   & 13th orb. & 14th orb. & 15th orb. & 16th orb. & 17th orb. & 18th orb. & 19th orb. & 20th orb. \\
\hline
SP1-min1  & 1.9814    & 1.9792    & 1.9764    & 1.9716    & 0.0317    & 0.0280    & 0.0253    & 0.0249 \\
SP1-sadd  & 1.9804    & 1.9784    & 1.9754    & 1.9711    & 0.0373    & 0.0298    & 0.0280    & 0.0239 \\
SP2-min2  & 1.9789    & 1.9772    & 1.9741    & 1.9737    & 0.0379    & 0.0286    & 0.0257    & 0.0226 \\
SP3-min1  & 1.9786    & 1.9773    & 1.9712    & 1.9694    & 0.0351    & 0.0319    & 0.0282    & 0.0272 \\
SP3-sadd  & 1.9800    & 1.9763    & 1.9703    & 1.9692    & 0.0385    & 0.0305    & 0.0299    & 0.0267 \\
SP4-sadd  & 1.9787    & 1.9769    & 1.9736    & 1.9691    & 0.0396    & 0.0303    & 0.0264    & 0.0217 \\
SP8-min1  & 1.9789    & 1.9772    & 1.9741    & 1.9737    & 0.0379    & 0.0286    & 0.0257    & 0.0226
\end{tabular}
\end{ruledtabular}
\end{table*}

\begin{figure}
\centering
\includegraphics[width=1.0\columnwidth]{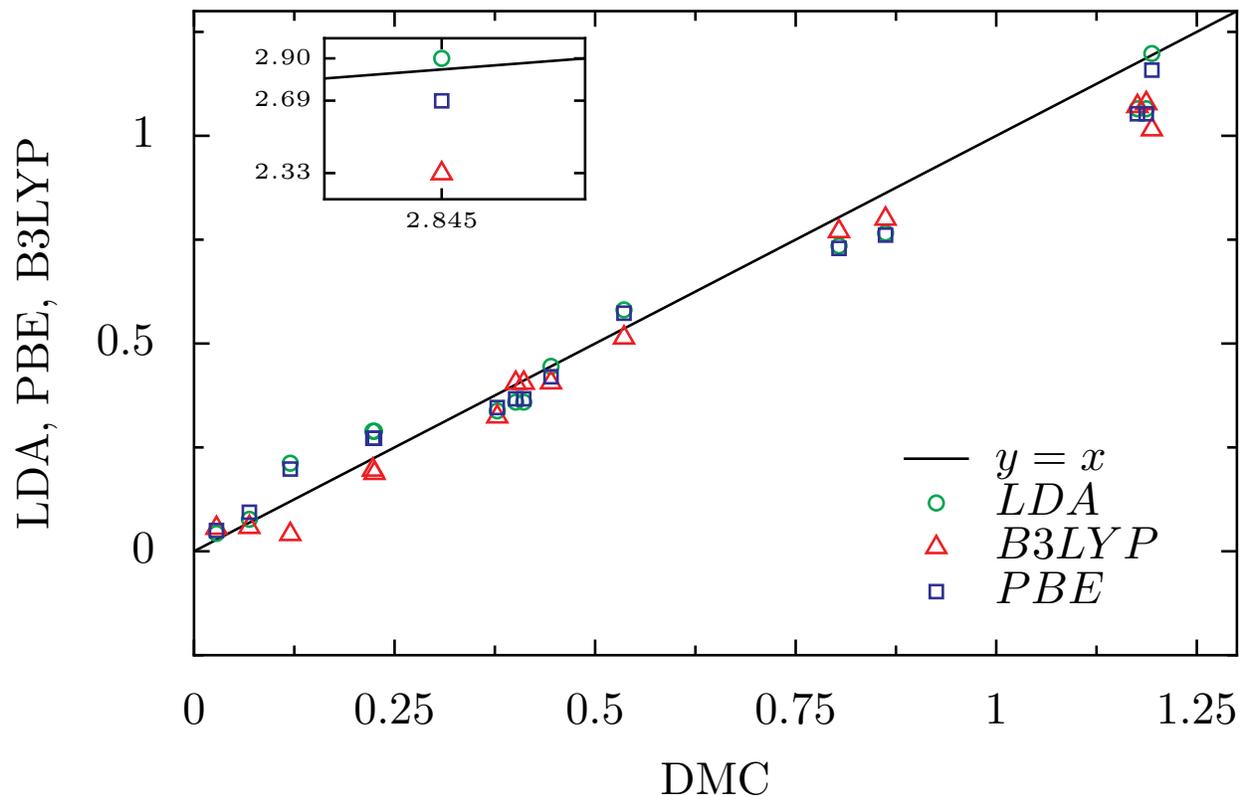}
\caption{(color online) The correlation of barrier heights from various DFT
functionals with the DMC barrier heights.
\label{fig:barriers}}
\end{figure}

Table~\ref{tab:barrierheights} also shows that the tight binding barrier heights are not reliable.
The situation is yet worse for the force fields and we have not even
attempted to give error bars. An additional complication, which will be discussed in next section, is that the potential energy
surfaces of force fields contain many fake minima and consequently also many fake saddle points connecting the
fake minima.

\section{Low Energy Configurations of Silicon Clusters and Crystalline Structures}

\subsection{Funnel-like structure of the PES}

The potential energy surface (PES) of the Si$_{16}$ cluster was explored
systematically with all of the aforementioned classical many-body potentials and the
Tight-Binding scheme using the minima hopping method~\cite{minhop} (MHM).
The MHM consists of a sequence of consecutive short molecular dynamics
runs and geometry relaxations. The MHM is a method that determines
the global minimum of a given PES as well as other low-lying energy configurations
very efficiently.

Atomic clusters range from structure seekers with well defined ground state configurations which can be found very rapidly with the MHM to glass-like systems for which it is very difficult
to find the ground state.
The speed with which a system finds its ground state is evidently a physical property
of the system and should carry over to most computational geometry-optimization algorithms.

However, we have found considerable differences in the speed of finding the ground state configuration with the MHM when using the various potentials to describe the Si$_{16}$ cluster.
Table \ref{tab:Minhopp} gives the average number $n_{MIN}$ of minima visited before finding the
global minimum. The differences in $n_{MIN}$ can be ascribed to the configurational density
of states (C-DOS) of the local minima for the particular potentials, discussed in
section~\ref{C-DOS}. A large $n_{MIN}$ indicates a high C-DOS in the low energy region and vice versa.
The Si$_{16}$ cluster for instance looks like a structure seeker with the Lenosky force field and
more like a glassy system with the Tersoff force field. In agreement with the observation in Fig.~\ref{fig:Plateau}
the MEAM (Modified Embedded Atom) ansatz~\cite{MEAM} of the Lenosky force field gives a relatively smooth potential energy surface.


\begin{table}[ht]
  \caption{Average values in 100 MHM runs. $n_{MIN}$ is the number of
minima visited before finding the ground state configuration.}
\begin{ruledtabular}
\begin{tabular} {l c }
    Method                & $n_{MIN}$  \\
    \hline
  EDIP                    &  85   \\
  Lenosky                 &  10   \\
  Reparametrized Lenosky  &  8    \\
  Tersoff                 & 116   \\
  Stillinger-Weber        &  31   \\
  Lenosky Tight-Binding   &  42   \\
  DFT                     &  32
\end{tabular}
\end{ruledtabular}
\label{tab:Minhopp}
\end{table}

\subsection{Ground state and low energy configurations of Si$_{16}$ isomers}
\begin{figure}            
\setlength{\unitlength}{1cm}
\subfigure[]{\includegraphics[width=0.48\columnwidth,angle=0]{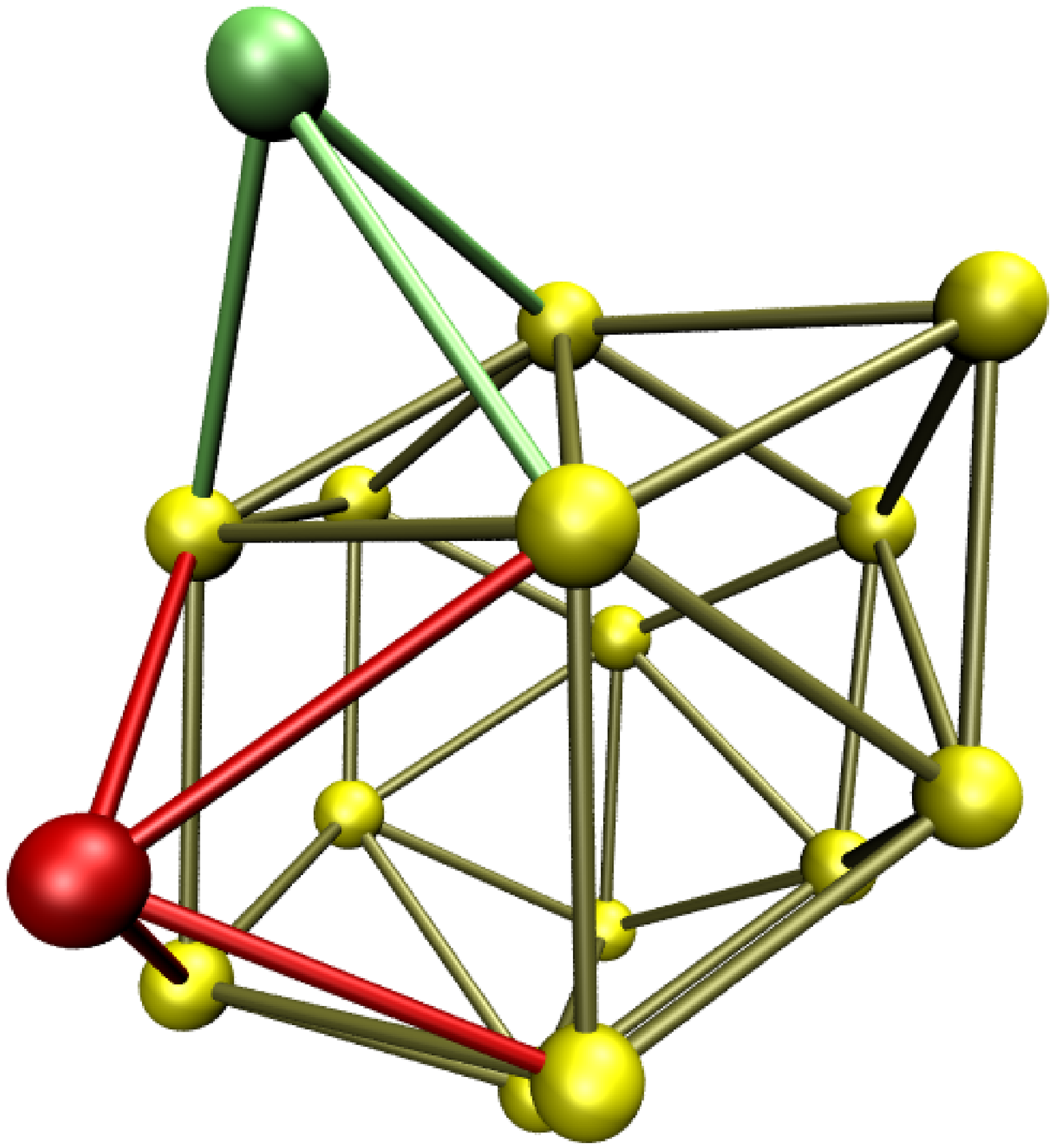}}  
\subfigure[]{\includegraphics[width=0.48\columnwidth,angle=0]{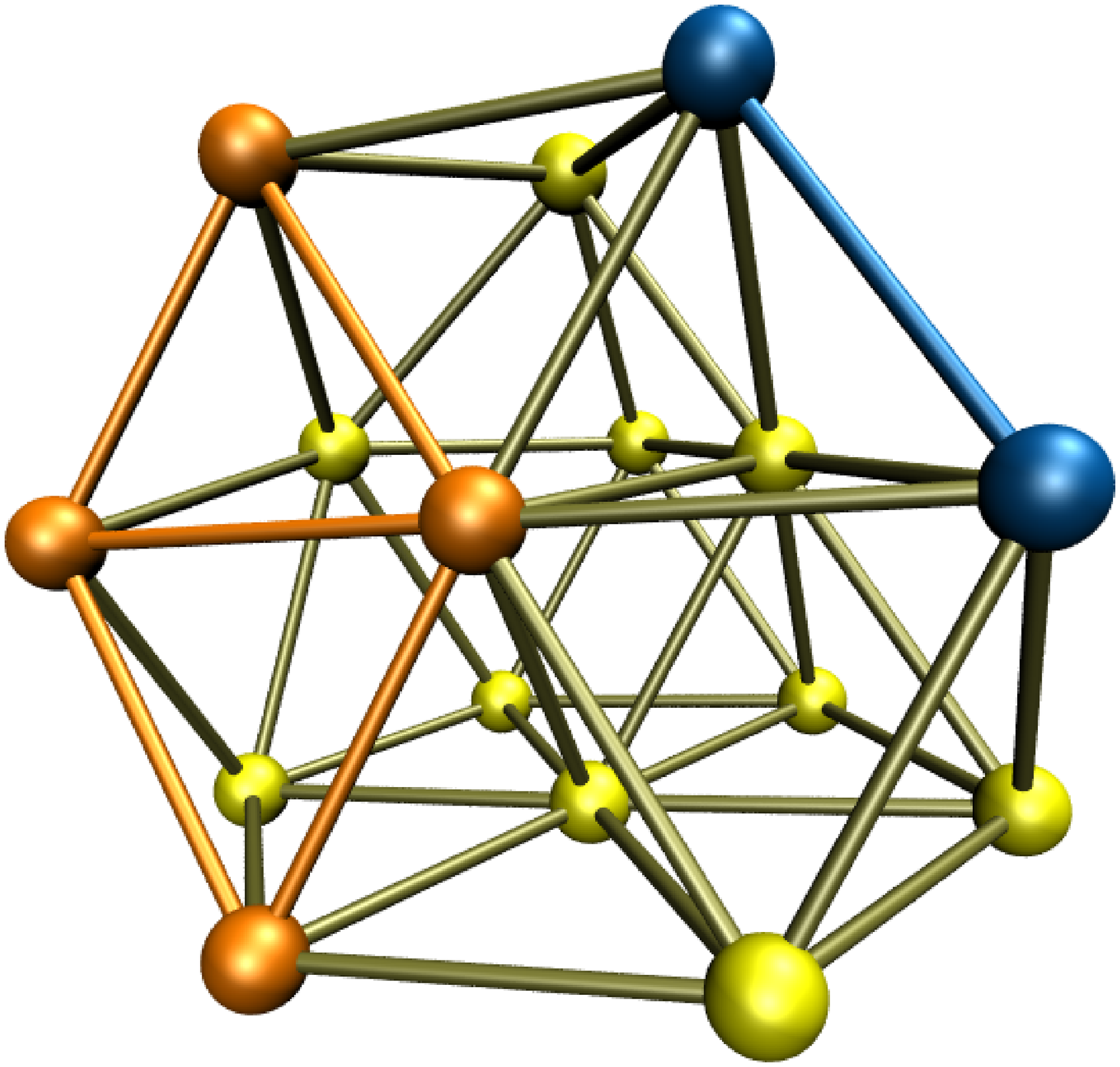}}  
\caption{\label{fig:Characteristic} (color online) (a) A surface atom forms a sharp
corner if it defines an
acute-angled tetrahedral (green) or pyramidal (red) structure together
with its three or four nearest neighbors respectively. (b) Two
neighboring surface atoms form a sharp edge if they make an
acute-angled tetrahedral structure with the two common nearest
neighbors (blue). Four neighboring surface atoms are part of a
facet if they form a plane (orange).}
\end{figure}

\begin{figure}            
\setlength{\unitlength}{1cm}
\subfigure[]{\includegraphics[width=0.47\columnwidth,angle=0]{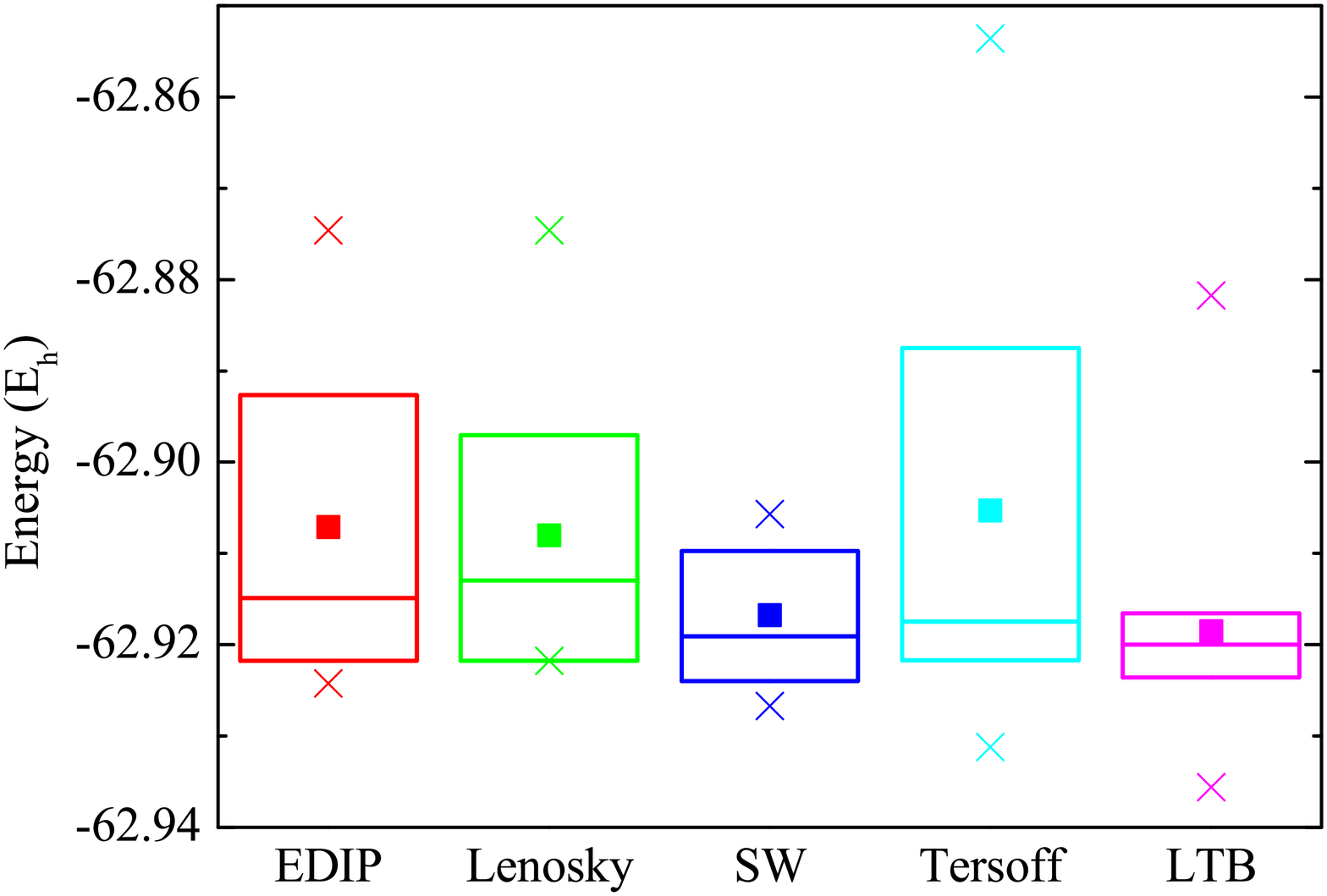}}
\subfigure[]{\includegraphics[width=0.47\columnwidth,angle=0]{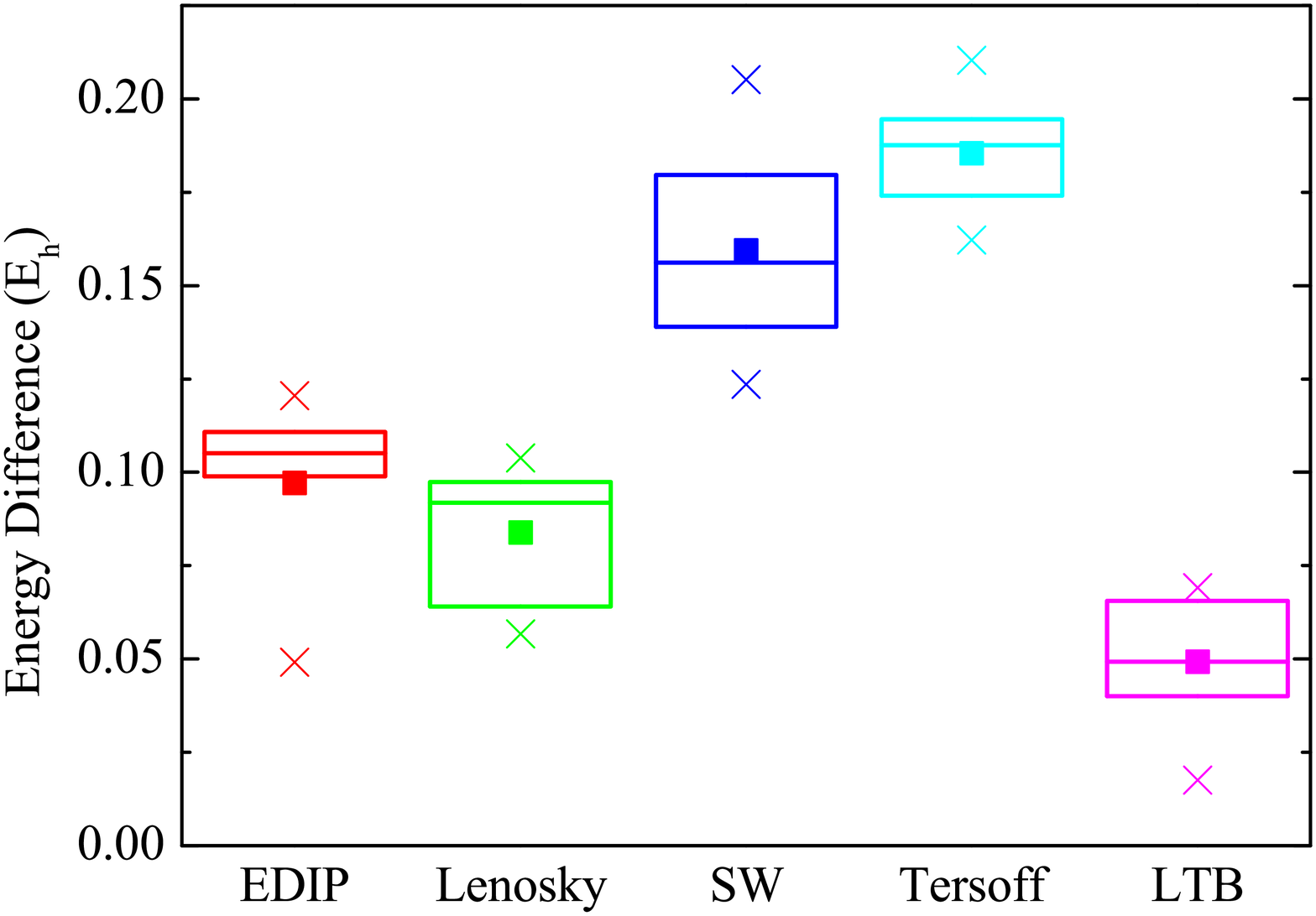}}
\subfigure[]{\includegraphics[width=0.47\columnwidth,angle=0]{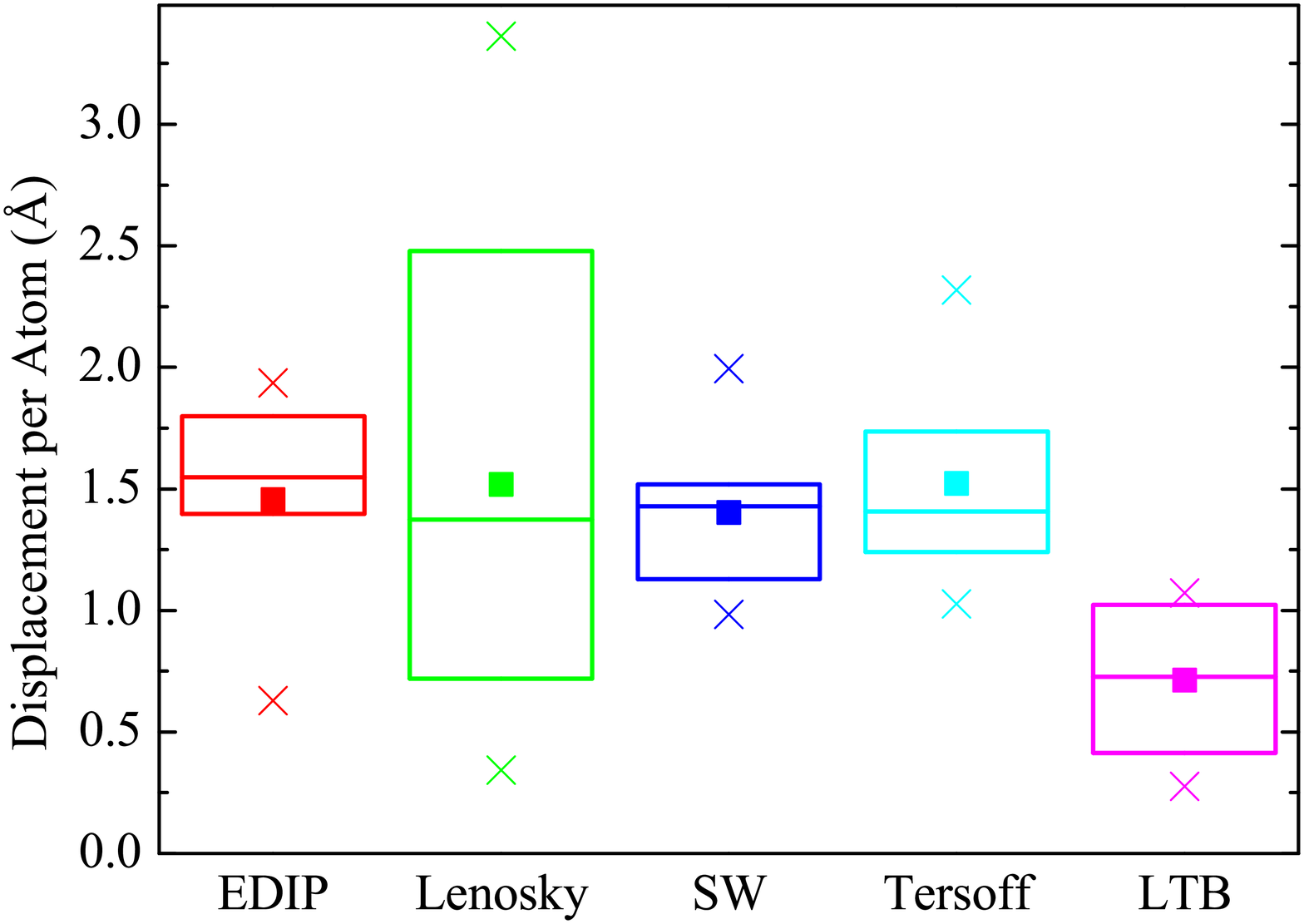}}
\caption{(color online) The box plots are based on ten low-lying structures of
Si$_{16}$. The boxes contain the values ranging from the lower to the
upper quartiles and the median is represented with horizontal lines. The
maximum and minimum as well as the mean values are plotted separately
with crosses and squares respectively. The absolute energy value found
after relaxation in DFT are represented in (a) with arbitrary origin.
Plot (b) shows the difference in DFT energy before and after relaxation.
Plot (c) shows the average displacements per atom before and after
relaxation. Unitary transformations of the initial and final structure
were performed to diagonalize the moment of inertia tensor with respect
to each atom. The transformations which resulted in the lowest
displacement were chosen for the plot. In plots (b) and (c) small values indicate
better agreement of the potential with DFT results.}
\label{fig:Boxplots}
\end{figure}

A database of stable configurations was generated by visiting 1000
different local minima with the MHM in each potential. To verify the
accuracy of the potentials the ten energetically lowest structures were
relaxed to the nearest local DFT minimum. The results of the investigation will be
discussed separately for each potential and are represented in Fig.~\ref{fig:Boxplots}.
The  geometrical features to characterize surface properties of Si$_{16}$ isomers are
described in Fig.~\ref{fig:Characteristic}.

\textbf{EDIP:}
The ground state configuration is a hollow oblate spheroid consisting
of four parallel planes. Each plane contains four atoms forming a
square which are rotated by 45\symbol{23} with respect to the neighboring
planes. The top and bottom squares have an edge length 2.49\AA{}, the
intermediate planes 4.27\AA{}. In this configuration all atoms are five
fold coordinated with an average bond length of 2.50\AA{}.
The first nine excited configurations have the same general features
as the ground state and only differ by forming or breaking of up to
three bonds. They cover an energy range of $0.5$~eV. When relaxed in DFT
most of the structures are heavily deformed. In general the void regions
within the structure collapse and the shape has a tendency to get
elongated. 7 structures show sharp corners and edges and two structures
exhibit extended facets. Only one minimum was found to be stable in DFT.

\textbf{Lenosky:}
The third and the fourth lowest energy configurations are oblate spheroid
and are the same as the second lowest and the global minimum configurations
with EDIP potential respectively.
However, the ground state
configuration with the Lenosky potential is highly spherical consisting
of four hexagonal curved panels with six-fold coordinated center atoms.
Although maintaining the general form, the relaxation in DFT reveals
that three of the four panels are transformed into planes. Remarkably,
an excited configuration was found to relax in DFT to the above mentioned ground
state configuration (energy difference in DFT $0.7$~eV). Furthermore,
there are four spherical and two elongated hollow structures, only one of
which is stable with DFT. All other excited states are deformed heavily
and the majority show both sharp corners and edges, covering an energy
range of $0.4$~eV.

\textbf{Stillinger-Weber:}
In contrast to the previous two potentials most configurations are
highly elongated and often contain pentagonal elements.
There are only three exceptions
including the ground state configuration which has a hollow elliptical
shape formed by 2 square and 8 pentagonal planes. Furthermore, none
of the geometries contains over-coordinated atoms. The structures
of all minima found with the Stillinger-Weber potential have a strong tendency to contain a large
number of sharp corners and only few facets when relaxed in
DFT. Although none of the structures are stable in DFT
the general elongated form is often
conserved and leads to low DFT energies, thus indicating an accurate
description of the low energy regions on the PES. The configurational
energies are scattered over a range of 0.80 eV.

\textbf{Tersoff:}
The ground state geometry of the Tersoff potential is identical
to the one found with the Stillinger-Weber potential. Only
one of the nine lowest energy configurations other than ground state
is elongated, the other eight are very
similar to the ground state with hollow spherical shapes.
The ninth excited state is $1.4$~eV above the ground state.
Similar to the Stillinger-Weber potential the structures do
not include over-coordinated atoms. All structures are deformed
after geometry optimizations with DFT and have a large number of sharp corners.
\begin{figure}[bh!]            
\setlength{\unitlength}{1cm}
\includegraphics[angle=-90,width=0.9\columnwidth]{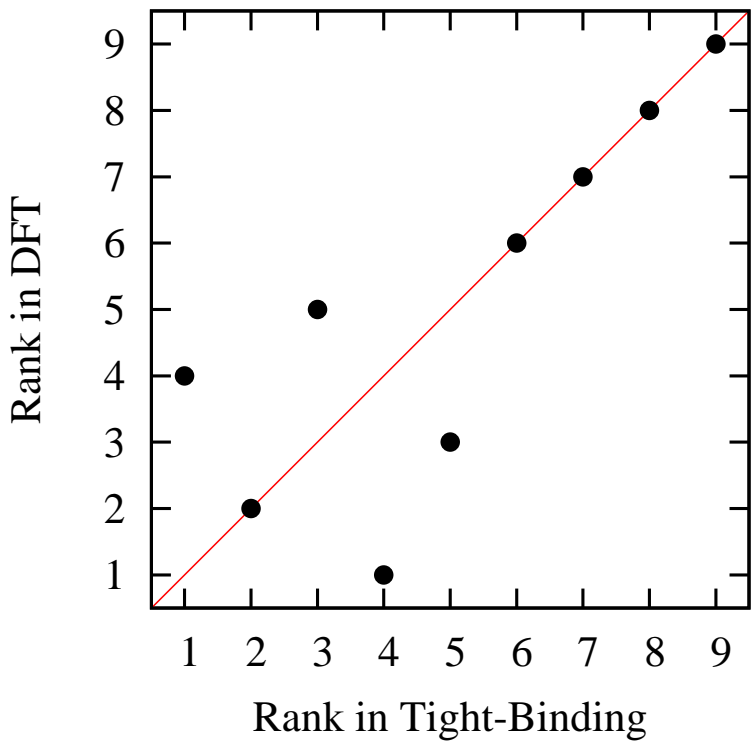}  
\caption{\label{fig:Ordering} (color online) Ordering of the local minima
energies in Lenosky tight-binding and DFT.}
\end{figure}

\textbf{Lenosky Tight-Binding:}
The Lenosky Tight-Binding scheme predicts a global minimum configuration
which is a slightly distorted Stillinger-Weber and Tersoff ground state
geometry. In contrast to the classical potentials both hollow elliptical
and elongated structures without void regions are predicted in equal
amounts. The energy of the ninth excited state lies only $0.15$~eV above
the ground state geometry, indicating an overall shallow PES. Although
some bond lengths are overestimated, all structures with only one exception
were found to be stable in DFT calculations. However, three configurations
with similar geometries converged to the same minimum structure. The ordering of
the minima with respect to the energies within the Tight-Binding scheme and the
DFT calculations is in fairly good agreement with the ideal
correlation (see Fig.~\ref{fig:Ordering}). While all four classical potentials fail to predict
stable low-lying Si$_{16}$ isomers the Lenosky Tight-Binding
scheme succeeds in most cases.

\begin{figure}            
\setlength{\unitlength}{1cm}
\subfigure[EDIP]{\includegraphics[width=0.48\columnwidth,angle=0]{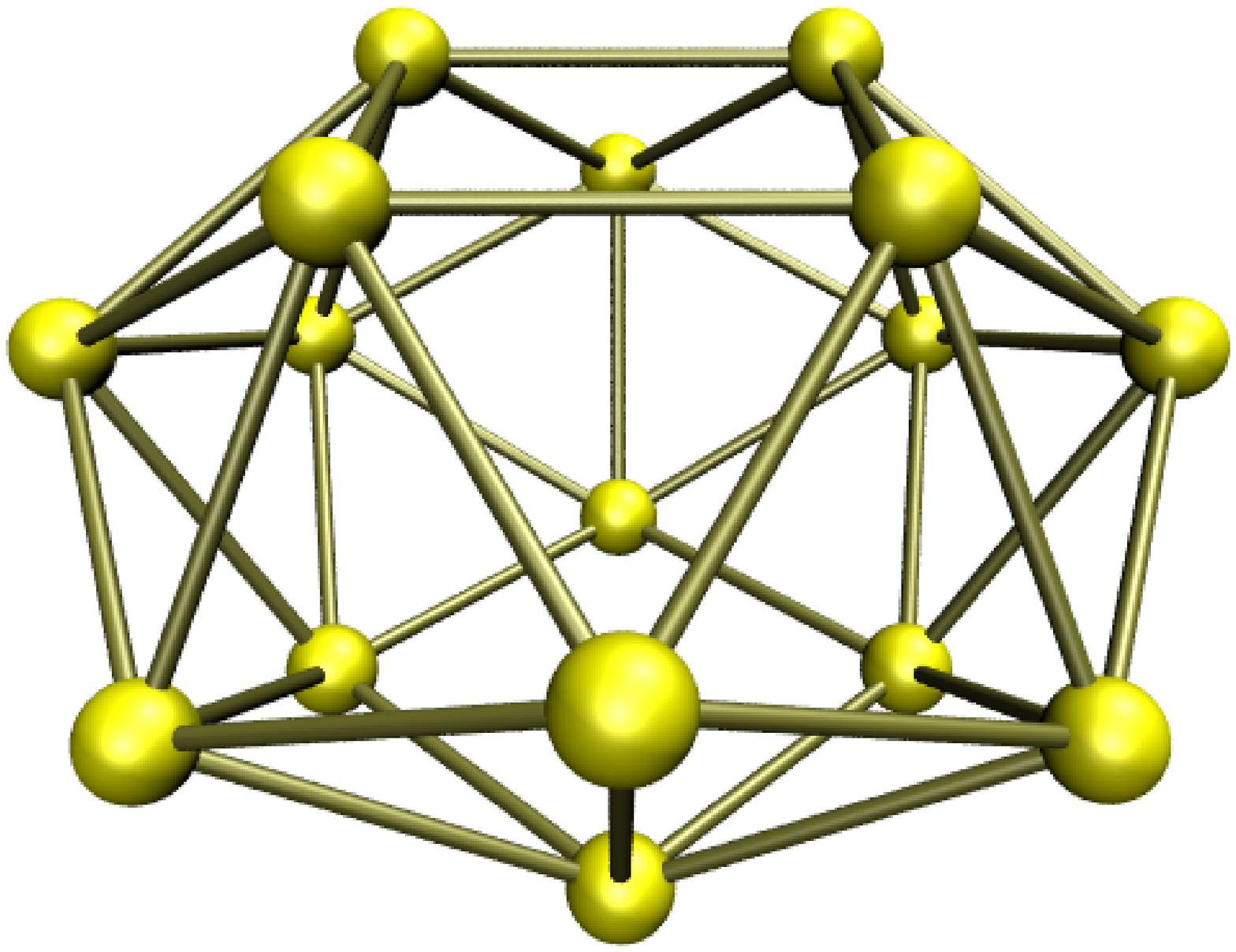}} 
\subfigure[Lenosky]{\includegraphics[width=0.48\columnwidth,angle=0]{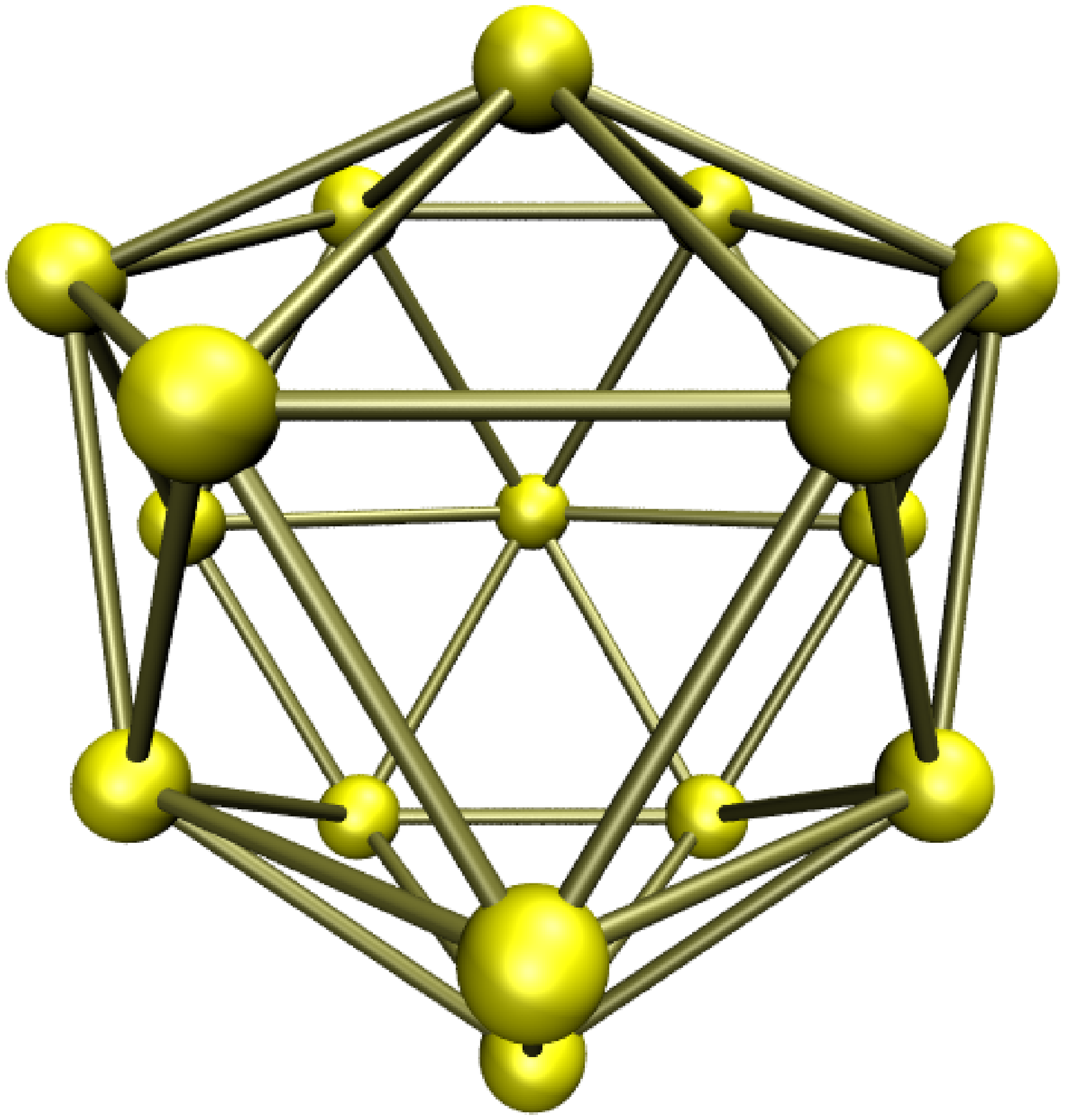}}  
\subfigure[Stillinger-Weber/Tersoff]{\includegraphics[width=0.48\columnwidth,angle=0]{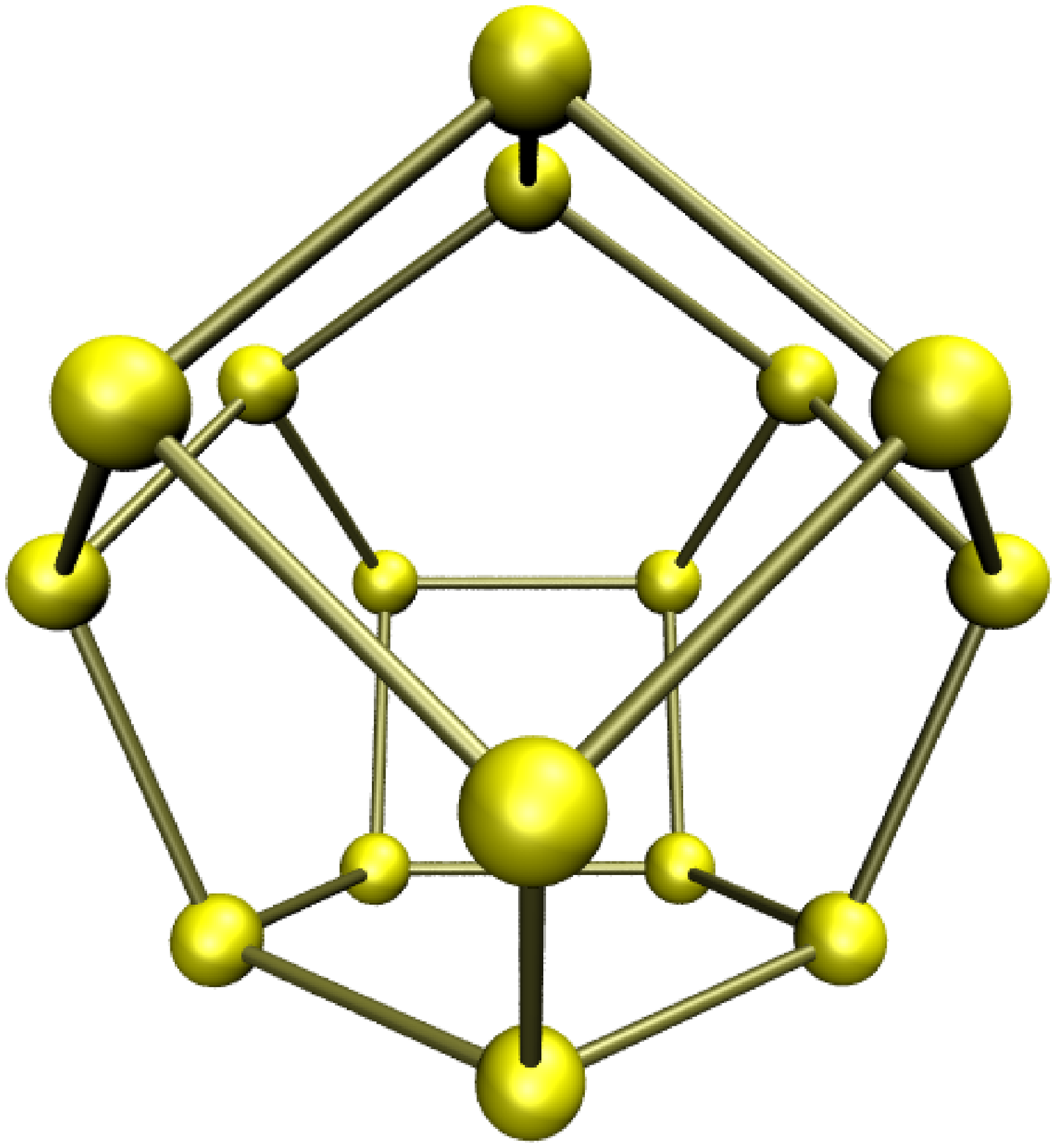}}  
\subfigure[Lenosky
Tight-Binding]{\includegraphics[width=0.48\columnwidth,angle=0]{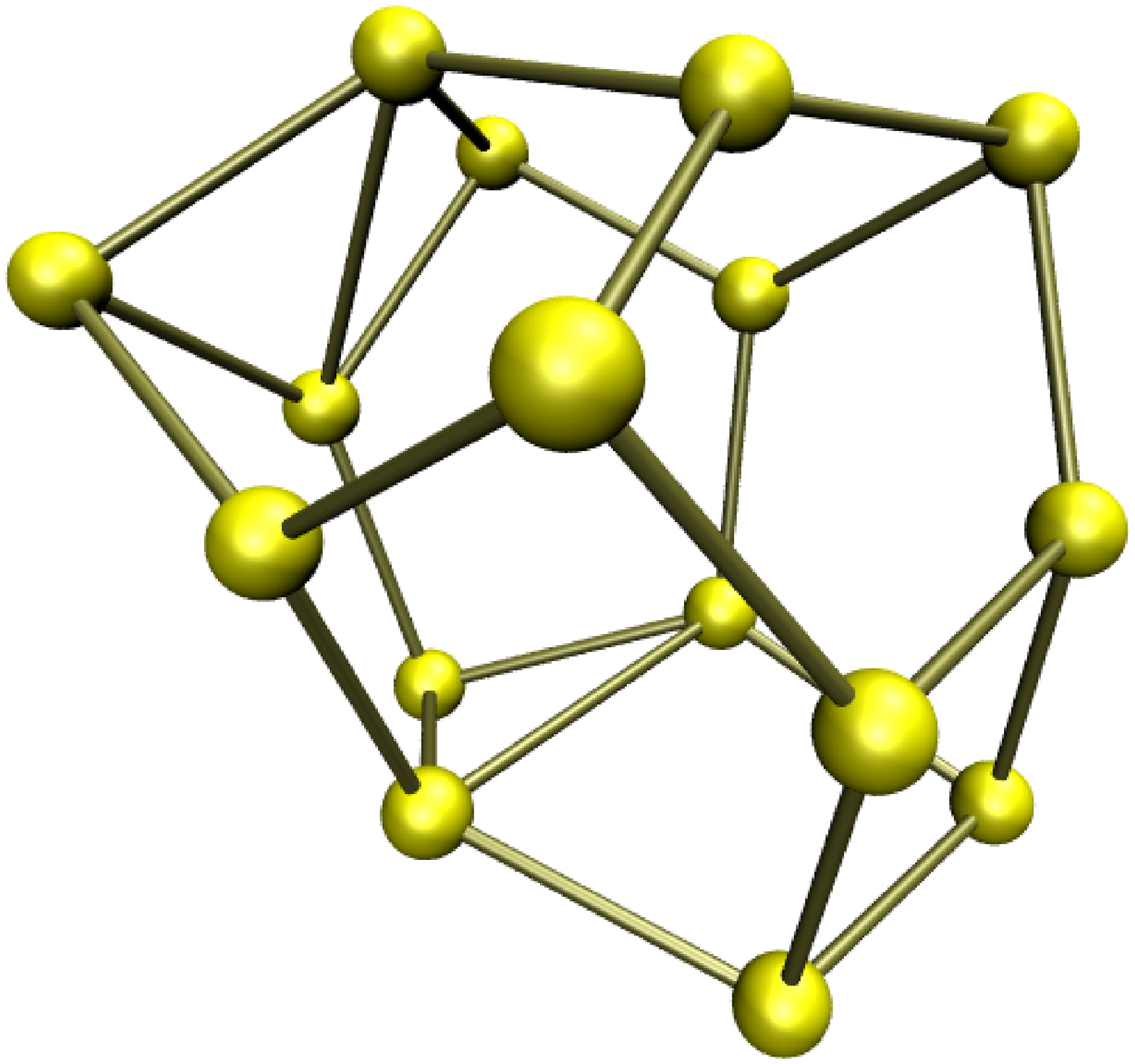}}
\caption{\label{fig:GlobalMinima} (color online) Global minimum configuration of
all five potentials found with the minima hopping method.}
\end{figure}

\subsection{Flat regions of the PES}
\begin{figure}[ht]             
\setlength{\unitlength}{1cm}
\includegraphics[angle=0,width=1\columnwidth]{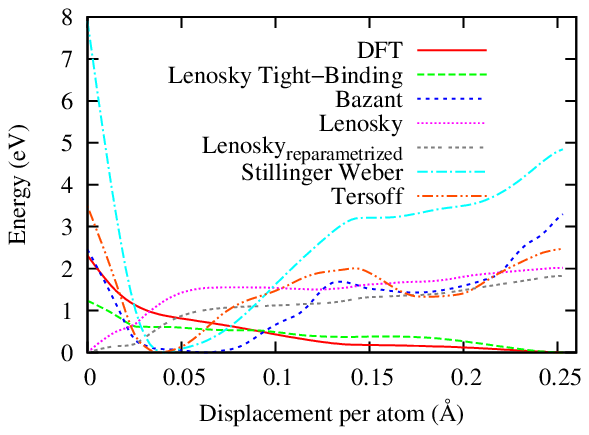}
\caption{\label{fig:Plateau} (color online) Energies of all potentials
along a relaxation path in a DFT
calculation plotted against the added up norm of the average atomic
displacement. The starting configuration is a Lenosky force field
local minimum. The energies are shifted such that the
minimum values are set to 0. $Lenosky_{reparametrized}$ is an
unpublished reparametrized Lenosky MEAM in which a
single FFCD is stable.}
\end{figure}

During DFT geometry relaxations one can encounter cases where the cluster
is distorted considerably even though the energy decreases only slightly.
Within these flat regions the norm of the force is small but may
increase while the monotonous downhill progress in energy is
preserved.
Many steps are necessary in the steepest descent DFT geometry relaxation to overcome
these flat plateau regions.
Only the Lenosky Tight-Binding scheme provides
an accurate energy trend when following the DFT relaxation pathway.
All classical potentials fail to even describe the lowering of the
configurational energy along the pathway (see Fig.~\ref{fig:Plateau}).
With the exception of the Lenosky force field these potentials
give a strongly oscillating  energy surface instead of a flat one
along the pathway. This is a first indication that the classical potentials
give a too rough PES. The MEAM ansatz of the Lenosky force field
seems to give smoother surfaces than the other classical potentials.

Furthermore, 120 random configurations were relaxed using
the different potentials and the largest eigenvalues of the
Hessian matrix were calculated when the local minimum configuration
was reached. The larger and highly scattered eigenvalues (Table~\ref{tab:Eval}) found
within the classical potentials indicate that these plateau-like
energy landscapes are poorly described by classical force fields.
The Stillinger-Weber potential provides the results closest to the
Lenosky Tight-Binding scheme, the most accurate among the potentials.
This is in agreement with the accurate overall description of low-lying
geometries with the Stillinger-Weber potential. The Tersoff potential
requires only a small number of relaxation steps, indicating a high
number of local minima and the absence of plateau regions on the PES.

\begin{table}
  \caption{Statistical data related to the Hessian matrix around local
minima of up to 120 random configurations (samples) of a Si$_{30}$
cluster. The second line contains the number of steps used in the
geometry relaxation, a combination of preconditioned steepest descent
and preconditioned DIIS method (except for the DFT). The corresponding
average, minimum and maximum of the largest and smallest eigenvalues of
the Hessian matrices are listed in the lines 3 to 8 (in eV/\AA{}$^2$).
The last line contains the average condition number $\kappa$ of the
corresponding Hessian matrices. $^*$Values only accurate in the first
decimal place.}
\begin{ruledtabular}
\begin{tabular}{l c c c c c c}


                               &   EDIP        &  Lenosky  &  SW
&   Tersoff  &   LTB     &     DFT$^*$\\
    \hline
samples                       &       120     &    119    &  120
&     119    &   119     &      40    \\
$\langle$steps$\rangle$       &       111.  &    99.  &  131.
&     39.  &   126.  &            \\
$\langle E_{large}\rangle$    &        61. &    43. &   38.
&     58. &    23. &      27. \\
$E_{large}^{min}$             &        33. &    30. &   32.
&     31. &    20. &      22. \\
$E_{large}^{max}$             &       116. &    95. &   44.
&    137. &    27. &      30. \\
$\langle E_{small}\rangle$    &         0.23 &     0.44 &    0.31
&      0.72 &     0.18 &       0.20 \\
$E_{small}^{min}$             &         0.004 &     0.062 &    0.056
&      0.15 &     0.003 &       0.03 \\
$E_{small}^{max}$             &         0.62 &     1.03 &    0.78
&      1.2 &     0.45 &       0.61 \\
$\langle \kappa \rangle$      &       476. &   132. &  175.
&     93. &   239. &     197.

  \end{tabular}
\end{ruledtabular}
  \label{tab:Eval}
\end{table}

\subsection{Defects in crystalline Silicon}

The MHM was used to explore the low energy region on the PES of
bulk silicon. Starting with crystalline cubic diamond structure
consisting of $216$ Si atoms, 200000 local minima were found
successively for each classical potential during the simulation.
For the Lenosky Tight-Binding scheme only 25000 structures could
be found due to limited computer time. Periodic boundary conditions
with respect to the ground state geometry were used to provide the
appropriate bulk conditions. The ten energetically lowest configurations
of each potential were used as input configurations for  geometry
relaxations in DFT.

\begin{table}
  \caption{The results of ten configurations of each potential
relaxed with DFT. The second column shows the number of stable
structures. The following columns show the number of structures
which relax to the bulk crystal, to a single FFCD or two FFCD which
are either neighboring (n-FFCD) or distant (d-FFCD).
}
\begin{ruledtabular}
\begin{tabular}{l r r r r r}
Method  &  Stable & Bulk & FFCD & n-FFCD & d-FFCD\\
\hline
EDIP    &      2  &   6  &   4  &     0  &    0\\
Lenosky &     10  &   1  &   0  &     2  &    7\\
SW      &      7  &   1  &   4  &     5  &    0\\
Tersoff &      2  &   4  &   6  &     0  &    0\\
LTB     &     10  &   1  &   1  &     4  &    4
 \end{tabular}
\end{ruledtabular}
  \label{tab:defects}
\end{table}

The correct ground state geometry, the well-known diamond structure, is predicted with all the potentials.
However, the structures of the first excited state of different force fields do not coincide.
For all potentials except the Lenosky force field it is a single four fold coordinate defect~\cite{FFCD} (FFCD).
The Lenosky potential on the other hand predicts a pair of two four fold
coordinated defects
\cite{FFCD} in different regions of the cell as the lowest energy defect structure.
The double FFCD is $3.99$~eV higher in energy compared to the diamond structure.

The majority of the eight other low energy geometries in
the EDIP potential are structures containing single
displaced atoms which are either four or five fold coordinated.
Similar structures can be found with the Tersoff potential.
All of these excited configurations are unstable in DFT calculations.
The Tersoff potential additionally
has minima at a variety of slightly distorted FFCDs which are unstable in DFT.

In contrast to the other three force fields, the Lenosky force field always
predicts pairs of FFCDs
as low-lying energy configurations. They are either neighboring and
share a common atom or are distant, i.e., located in different regions of the cell.
Even though the single FFCD, which must be the first excited state, is
not predicted by the Lenosky force field~\cite{note}, all other low-lying
energy configurations from the second to ninth excited states
are stable in DFT geometry optimization. Nonetheless, the sequence with respect to
the energy does not coincide with the sequence in DFT energies. The
Stillinger-Weber force field behaves very similarly. Only three
structures were found to be unstable, the other five excited states
all contain two interacting FFCDs.

The best accuracy can be found with the Lenosky Tight-Binding scheme.
All structures exist on the DFT Born-Oppenheimer surface
and the energy sequence is correctly described with the exception of
the 9$^{th}$ and 10$^{th}$ excited states. They are exchanged in
sequence and show an energy difference of $0.02$~eV with the Lenosky
Tight-Binding scheme and $0.04$~eV when calculated with DFT.

\begin{figure}            
\includegraphics[clip,angle=-90,width=1\columnwidth]{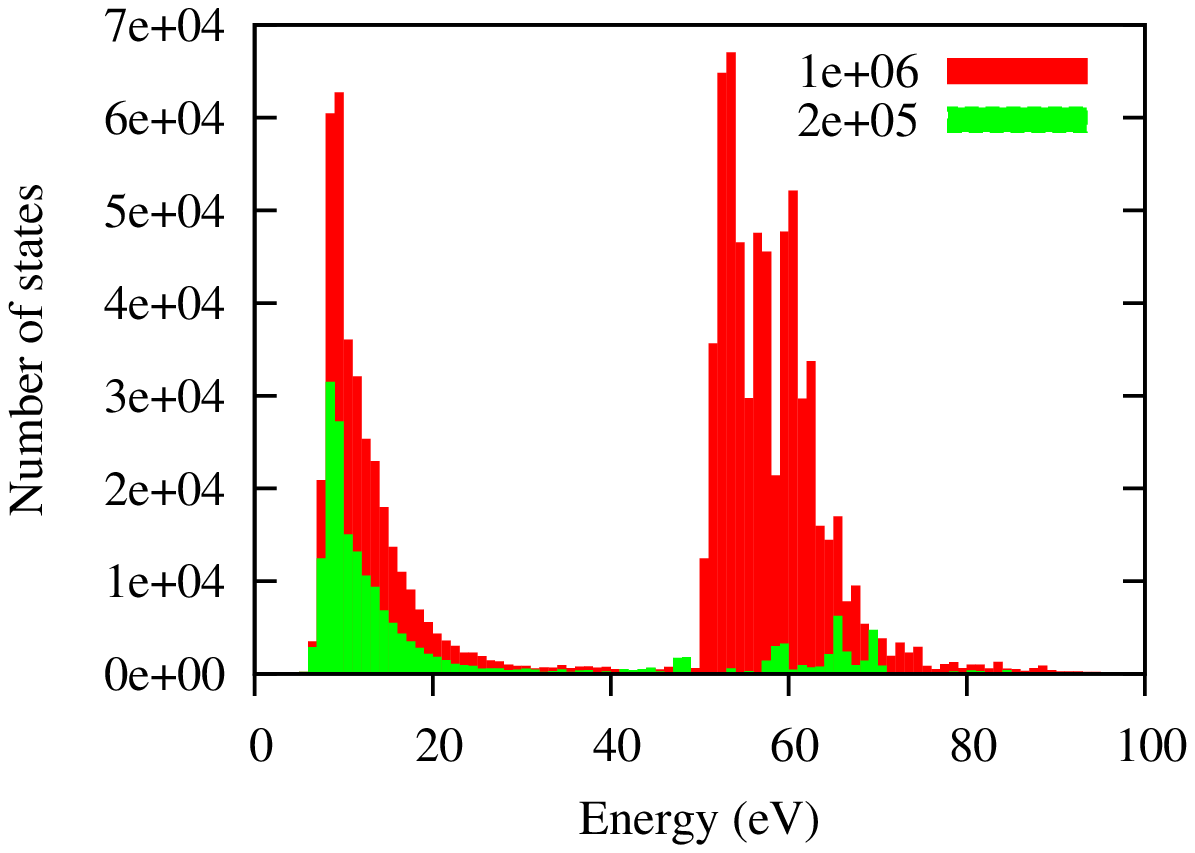}  
\caption{\label{fig:Convergence} (color online) The minima-hopping density-of-states, MH-DOS, represented by a
histogram with 100 bins. $1 \times 10^6$ and $2 \times 10^5$ structures were found with the
MHM, colored red and green respectively.}
\end{figure}

\begin{figure*}            
\setlength{\unitlength}{1cm}
\subfigure{\includegraphics[clip,width=0.05\columnwidth]{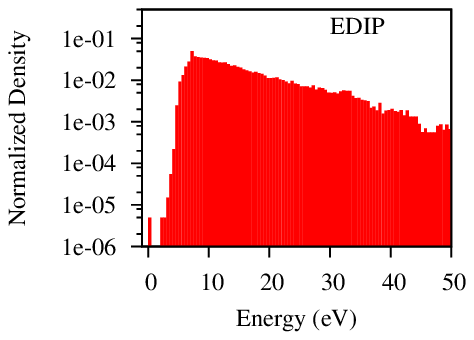}}  
\subfigure{\includegraphics[clip,width=0.30\columnwidth]{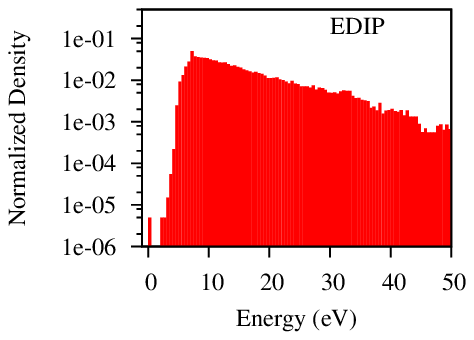}}  
\subfigure{\includegraphics[clip,width=0.30\columnwidth]{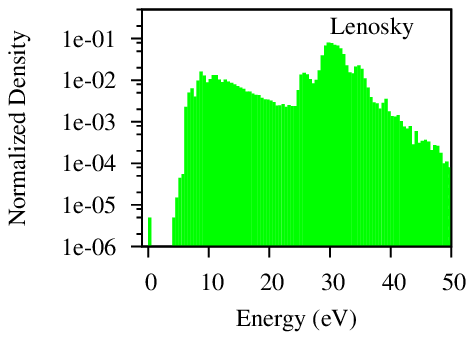}}  
\subfigure{\includegraphics[clip,width=0.30\columnwidth]{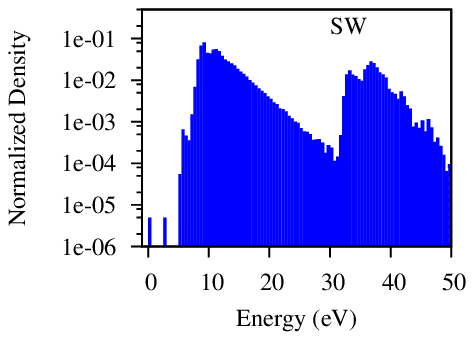}}  
\subfigure{\includegraphics[clip,width=0.05\columnwidth]{figs/ylabel.eps}}  
\subfigure{\includegraphics[clip,width=0.30\columnwidth]{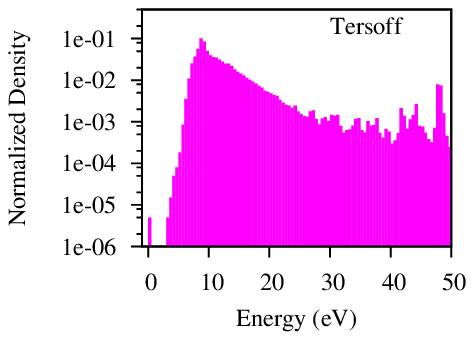}}  
\subfigure{\includegraphics[clip,width=0.30\columnwidth]{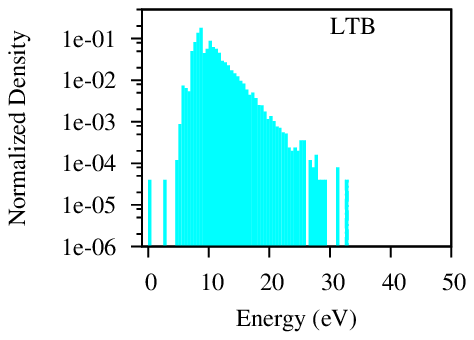}}  
\caption{\label{fig:density} (color online) The normalized MH-DOS
as represented by a histogram consisting of 100 bins on a logarithmic
scale. While 200000 values were used for the classical potentials only 25000 could
be calculated with the Lenosky Tight-Binding scheme. The energy is shifted such that
the ground state has energy 0.}
\end{figure*}

\subsection{Configurational density of states of local minima\label{C-DOS}}

To describe the overall characteristics of the potential energy surface
we chose the configurational density of states (C-DOS), i.e.,
the number of configurations per energy interval. We approximate this C-DOS by the
minima hopping density of states (MH-DOS) which
is obtained simply by sampling the low energy region with the
MHM and counting the number of distinct minima found in an energy
interval. It has to be stressed that, in the plots we present in this
paper, a more or less complete sampling of all minima can only be achieved
in a very small interval around the global minimum. Only in this
small interval of several eV we observe in the MH-DOS the expected exponential
growth of the number of local minima with respect to the energy of the C-DOS.
In our plots we show however a much larger energy interval
where the number of states is the true number of states multiplied by the probability
that a configuration in this energy range will be visited. Since this
probability decreases with increasing energy the MH-DOS tends to
zero for large energies in all our plots whereas the C-DOS would be orders of magnitude larger.
Since the minima hopping method maps out higher and higher energy configurations
when the minima hopping run is allowed to continue longer and longer, we can
can compare the results of our standard length MHM run, where 200 000 configurations were found,
to the results of a longer run, where 1000000 configurations were found.
As one can see from Fig.~\ref{fig:Convergence}
the MH-DOS and C-DOS agree only within the first few bins of the exponential growth region.
The lowest energy minima correspond to point defects. The onset of the exponential growth
region is due do a growing number of defects (mainly of the FFCD type)
which lead continuously to amorphous structures.
The longer simulation in Fig.~\ref{fig:Convergence} also shows
a second peak at around 55 eV. This peak is due to amorphous configurations which are related
to a sheared crystalline structure. Since we do not relax the simulation cell sheared structures
cannot relax.

The reason why we show the MH-DOS over an energy interval which is much
larger than the interval within which we can obtain a reliable C-DOS is the following.
If there were good agreement between the C-DOS obtained from different force fields
the MH-DOS would also agree. As seen from Fig.~\ref{fig:density} the MH-DOS
obtained from different force fields are drastically different and one can therefore conclude
that the C-DOS are also drastically different.
The MH-DOS of all potentials are included in Fig.~\ref{fig:density}.
Stillinger-Weber and the Lenosky Tight-Binding show similar
features in the low energy region, e.g., the energy gap between the
single FFCD and higher excited states and the spike
between 7 and 8 eV. While the EDIP and Tersoff potential
show only a single major peak around 10 eV, both Lenosky and
Stillinger-Weber have a second peak located at about $35$~eV
which corresponds, as discussed above, to sheared structures.
This is due to the fact that for these two potentials the C-DOS of
unsheared amorphous structures is much lower than for the other potentials
and the MHM starts therefore sampling higher energy regions corresponding to
sheared structures. The differences in the C-DOS are responsible for the
different speeds with which the global minimum is
found (see Table \ref{tab:Minhopp}).

\section{Conclusions}
We have shown that DFT and in particular LDA barrier heights are rather accurate for
rearrangement processes occurring in silicon clusters. This is good news since the estimation
of diffusion coefficients and other dynamical properties in silicon systems are frequently based on
DFT calculations. Since it is well established that DFT schemes give highly accurate results
for structural properties, i.e., local minima of the potential energy surface, DFT calculations
are able to provide very reliable potential energy surfaces for silicon systems.
The bad news is that force fields, that are widely used for dynamical
simulations in large silicon systems, do not faithfully describe the potential energy surface.
With the exception of the MEAM based Lenosky force field, all force fields give rise to
potential energy surfaces that are too rugged.
In an extended crystalline environment
most force fields greatly overestimate the configurational density of states
because they give rise to many fake defect structures
which do not exist in more accurate schemes.  The situation is even worse
for non-periodic systems where the large majority of stable structures are fake.
Simulations based on the use of a single force field should therefore be viewed with caution
and should be verified by density functional calculations whenever this is feasible.

\section{Acknowledgments}
We thank Gustavo Scuseria for interesting discussions. Financial support was provided by the
Swiss National Science foundations. The calculations were done at the Swiss National Supercomputing
center (CSCS) in Manno.
The work at Cornell University was supported by the National Science
Foundation under grants EAR-0703226 and EAR-0530813.  The research used computational
resources provided by the National Energy Research Scientific
Computing Center, which is supported by the Office of Science of the
U.S. Department of Energy, by Teragrid at the National Center for
Supercomputing Applications, which is supported by the National
Science Foundation and by the Computational Center for Nanotechnology
Innovations, which is supported by the state of New York.

An implementation with an identical calling sequence of all the force fields used in this work
can be downloaded from www.unibas.ch/comphys/comphys.

\clearpage

\bibliography{siliconlandscape}

\end{document}